\documentclass[conference]{IEEEtran}

\usepackage{times}
\usepackage{epsfig}
\usepackage{graphicx}
\usepackage{amsmath}
\usepackage{amssymb}
\usepackage{algorithmic}
\usepackage{adjustbox}
\usepackage{booktabs}
\usepackage{comment}
\usepackage{acronym}
\usepackage{xspace}
\usepackage{amsfonts} 
\usepackage[mathscr]{eucal}
\usepackage{xcolor}
\usepackage{adjustbox}
\usepackage{booktabs}
\usepackage{xcolor}     
\usepackage{amsmath}
\usepackage{amssymb}
\usepackage{graphicx}
\usepackage{graphics}
\usepackage{acronym}
\usepackage{xspace}
\usepackage{hyperref}
\usepackage{url}
\usepackage{xcolor}
\usepackage{array,multirow,graphicx}
\usepackage{epsfig}
\usepackage{subcaption}
\usepackage{longtable}
\usepackage{media9}
\usepackage{lscape}
\usepackage{slashed}
\usepackage{amsmath}
\usepackage{textcomp}
\usepackage{amssymb}
\usepackage{cancel}
\usepackage{tikz}
\usetikzlibrary{tikzmark,shapes.misc}
\usepackage{mathtools}
\newcommand\todo[1]{\textcolor{blue}{TODO: #1}}
\newcommand\tocite[1]{\textcolor{blue}{[REFERENCE]}}
\usepackage{acronym}
\usepackage{xspace}
\usepackage{bbding}
\usepackage{comment}
\usepackage{afterpage}
\usepackage{adjustbox,lipsum}

\acrodef{CNN} 		[\textsc{CNN\xspace}]				{Convolutional Neural Network}

\usepackage{subcaption}

\acrodef{ML} 		[\textsc{ML\xspace}]				{Machine Learning}
\acrodef{DL} 		[\textsc{DL\xspace}]				{Deep Learning}
\acrodef{ANN} 		[\textsc{ANN\xspace}]				{Artificial Neural Network}
\acrodef{DNN} 		[\textsc{DNN\xspace}]				{Deep Neural Network}
\acrodef{CNNs} 		[\textsc{CNNs\xspace}]				{Convolutional Neural Networks}
\acrodef{CNN} 		[\textsc{CNN\xspace}]				{Convolutional Neural Network}
\acrodef{MTL} 		[\textsc{MTL\xspace}]				{Multi-Task Learning}
\acrodef{RL} 		[\textsc{RL\xspace}]				{Reinforcement Learning}
\acrodef{LL} 		[\textsc{LL\xspace}]				{Lifelong Learning}
\acrodef{NLP} 		[\textsc{NLP\xspace}]				{Natural Language Processing}
\acrodef{LSTM} 		[\textsc{LSTM\xspace}]				{Long Short-Term Memory}
\acrodef{MAML} 		[\textsc{MAML\xspace}]				{Model Agnostic Meta Learning}
\acrodef{MMMTL} 	[\textsc{3MTL\xspace}]	            {Multi-modal Multi-task Meta Transfer Learning}
\acrodef{ADME}  	[\textsc{ADME\xspace}]	            {Absorption, Distribution, Metabolism, and Excretion}
\acrodef{GAN}   	[\textsc{GANs\xspace}]	               {Generative Adversarial Nets}
\acrodef{VAE}   	[\textsc{VAEs\xspace}]	               {Variational Auto-Encoders}
\acrodef{MTML}   	[\textsc{MTML\xspace}]	               {Multi-task Meta Learning}

\begin{document}

\title{Magnification Prior: A Self-Supervised Method for Learning Representations on Breast Cancer Histopathological Images}

\makeatletter

\makeatother

\author{
    \IEEEauthorblockN{
        Prakash Chandra Chhipa\textsuperscript{1,*},
        Richa Upadhyay\textsuperscript{1},
        Gustav Grund Pihlgren\textsuperscript{1},
        Rajkumar Saini\textsuperscript{1},\\
        Seiichi Uchida\textsuperscript{2} and
        Marcus Liwicki\textsuperscript{1}\\
        \textit{\textsuperscript{1}Machine Learning Group, EISLAB, Lule\aa~Tekniska Universitet, Lule\r{a}, Sweden}\\
        \textit{\{first.middle.last\}@ltu.se}\\
        \textit{\textsuperscript{2}Human Interface Laboratory, Kyushu University, Fukuoka, Japan}\\
        \textit{uchida@ait.kyushu-u.ac.jp}\\
        \textit{\textsuperscript{*}Corresponding author - prakash.chandra.chhipa@ltu.se}
     }
    }

\maketitle

\begin{abstract}

This work presents a novel self-supervised pre-training method to learn efficient representations without labels on histopathology medical images utilizing magnification factors.
Other state-of-the-art works mainly focus on fully supervised learning approaches that rely heavily on human annotations.
However, the scarcity of labeled and unlabeled data is a long-standing challenge in histopathology. Currently, representation learning without labels remains unexplored in the histopathology domain.
The proposed method, Magnification Prior Contrastive Similarity (MPCS), enables self-supervised learning of representations without labels on small-scale breast cancer dataset BreakHis by exploiting magnification factor, inductive transfer, and reducing human prior.
The proposed method matches fully supervised learning state-of-the-art performance in malignancy classification when only 20\% of labels are used in fine-tuning and outperform previous works in fully supervised learning settings for three public breast cancer datasets, including BreakHis. Further, It provides initial support for a hypothesis that reducing human-prior leads to efficient representation learning in self-supervision, which will need further investigation.
The implementation of this work is made available online on GitHub\footnote{\scriptsize{\url{https://github.com/prakashchhipa/Magnification-Prior-Self-Supervised-Method}}}.

\textit{Keywords: self-supervised learning, contrastive learning,  representation learning, breast cancer, histopathological images, transfer learning, medical images}
\end{abstract}
\section{Introduction}

\begin{figure}[!ht]
    \centering
    \includegraphics[width =\linewidth]{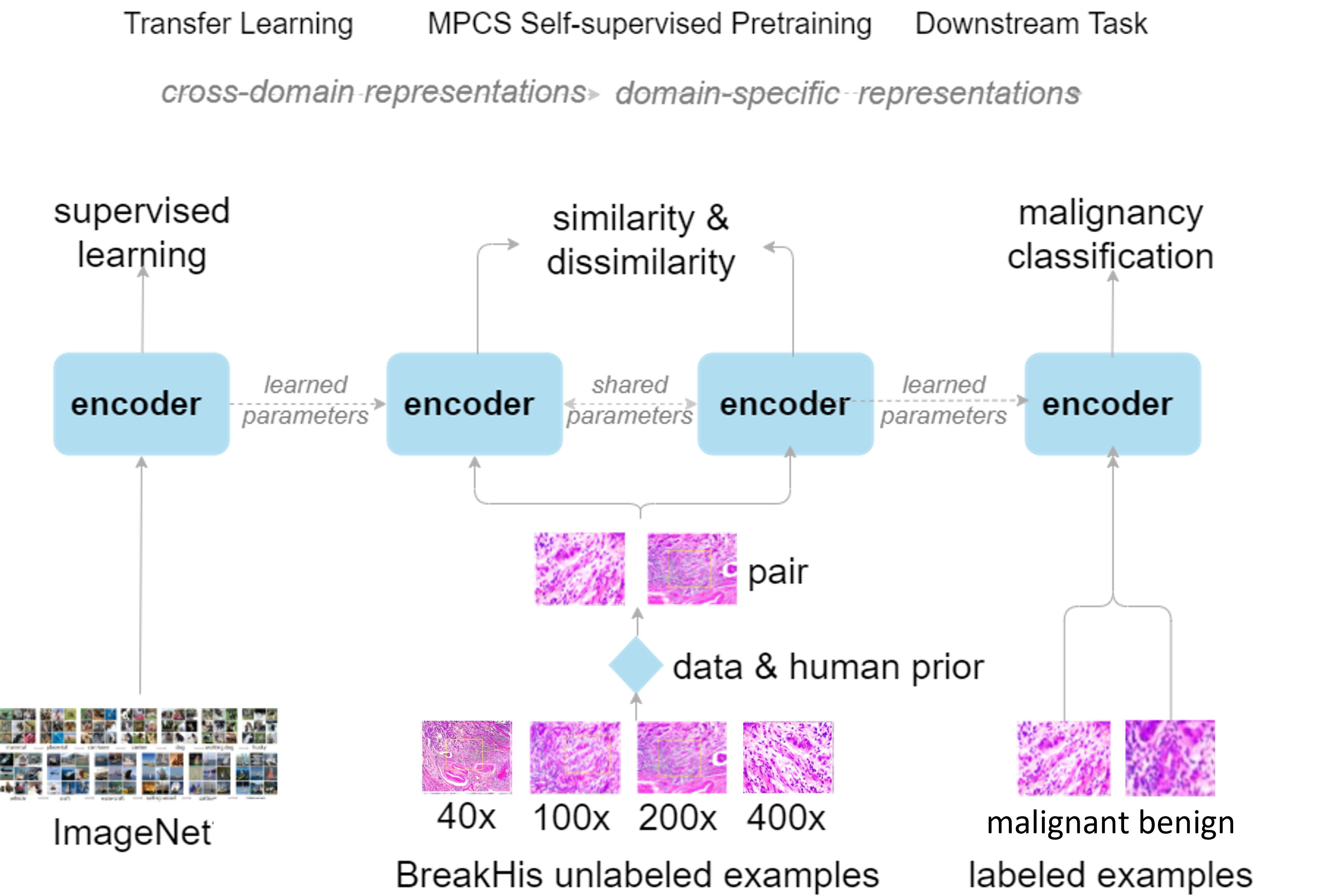}
    \vspace{-0.6cm}
    \caption{The proposed approach comprises three steps: (1)~Parameters initialization with supervised ImageNet weights. (2)~Self-supervised pre-training on unlabeled BreakHis histopathology images using proposed method \textbf{Magnification Prior Contrastive Similarity} to provide positive pairs by exploiting supervision signal, e.g., magnification from data and reducing human prior. (3) Fine-tuning on histopathology images for the downstream task.}
    \label{fig:mpcs_intro}
    \vspace{-6mm}
\end{figure}

Cancer diagnosis by analyzing histopathological whole-slide images (WSI) is an active research field in machine learning~\cite{komura2018machine}.


A challenge for the supervised learning approaches applied to histopathological WSI is the scarcity of labeled data.
Furthermore, label information for digital WSI is also limited and does not provide details of the affected region at different magnifications, as in dataset BreakHis~\cite{spanhol2016dataset}.
Representations learned through supervised learning might suffer as such methods typically require a large amount of labeled data.
This can lead to sub-optimal performance on downstream tasks.

Exploring efficient representation learning on small-scale histopathological WSI data using objectives that do not require labels is a promising approach as it requires a lower amount of labeled data needed to learn successful downstream models.


This work proposes a novel self-supervised learning (SSL) method based on contrastive joint embedding called Magnification Prior Contrastive Similarity (MPCS), to learn efficient representations without labels.
The proposed method uses a magnification factor (a signal from data) to construct positive pairs for contrastive similarity.
MPCS uses magnification factors to enable SSL on the small-scale dataset. This work also hypothesizes that reducing human inducted prior in SSL methods enhances representation learning. 

The proposed method MPCS conducts self-supervised pre-training on histopathological WSI on the BreakHis~\cite{spanhol2016dataset} dataset with two different backbone encoders Efficient-net~\cite{tan2019efficientnet} and dilated ResNet-50~\cite{yu2017dilated}, pre-trained on ImageNet~\cite{deng2009imagenet}.
The effectiveness of the learned representations is assessed by fine-tuning for the downstream task of malignancy classification on multiple public datasets e.g. BACH~\cite{aresta2019bach}, Breast Cancer Cell Dataset~\cite{gelasca2008evaluation} including BreakHis dataset.
The complete approach is depicted in Figure~\ref{fig:mpcs_intro}.
Following are the main contributions: 
\begin{enumerate}
\item Enables self-supervised learning on small-scale histopathology dataset by exploiting a data prior
\item Hypothesizes a relation between human inducted prior and data prior for self-supervised representation learning
\end{enumerate}
With contributions mentioned above, this work demonstrates significantly improved performance on downstream tasks on three public datasets. It provides preliminary empirical support that 1) reducing human-prior leads to efficient representation learning and 2) learning being magnification invariant by cross-magnification evaluation.

\section{Related Work} \label{related_work}
Most efforts in machine learning for histopathological analysis have been using supervised learning.
However, current supervised learning methods struggle when labeled data is scarce~\cite{komura2018machine}.
Other methods, e.g. pseudo-label and transfer learning, are often used with supervised learning to make up for the lack of labeled data.
Examples of such methods are augmentation~\cite{lin2018scannet} for the first category and feature-extraction and selection~\cite{spanhol2017deep, wang2020breast} for the second.

One public histopathological image dataset that poses such a challenge of data- and label scarcity is BreakHis~\cite{spanhol2016dataset}.
Many different methods have been applied to the BreakHis~\cite{spanhol2016dataset} dataset, most of which utilize a \ac{CNN} in conjunction with one or both of the methods for handling data scarcity above.

Table~\ref{tab:previous_work} contains a summary of a few such approaches along with some of the strategies used on BreakHis~\cite{spanhol2016dataset}.
In the table, custom means a specialized or novel method introduced. Simple augmentations refer to one or more operations among rotation, flipping, cropping, shifting, and zooming.
Another breast cancer dataset is BACH~\cite{aresta2019bach}, on which previous works in transfer learning TL~\cite{vesal2018classification}, patch-based PT~\cite{roy2019patch}, and hybrid networks HN~\cite{yan2020breast} shows performance improvements. Further, the small-scale dataset Breast Cancer Cell Dataset~\cite{gelasca2008evaluation} is also being evaluated for binary tasks in various previous works based on shearlet transform ST~\cite{rezaeilouyeh2016microscopic} and attention methods are ATN~\cite{ilse2018attention} and MATN~\cite{konstantinov2022multi}.


\begin{table}[t]
\caption{Methodologies Used on the BreakHis Dataset.}
\vspace{-0.3cm}
\label{tab:previous_work}
\resizebox{\columnwidth}{!}{%
\begin{tabular}{c|ccccc}
\hline
    Work &  Model & Augmented & Extra Training &   Ensemble &  Evaluation\\
\hline
    Deep~\cite{spanhol2016breast} & AlexNet~\cite{krizhevsky2012imagenet} variant & No & No & No & 5 trials\\
    MI~\cite{bayramoglu2016deep} & Custom CNN & Simple & Custom & No & 5-fold\\
    GLPB~\cite{ataky2020data} & TCNN~\cite{andrearczyk2016using} & Custom & No & No & 5-fold\\
    MIL~\cite{sudharshan2019multiple} & Various & MIL~\cite{keeler1991integrated} & No & No & 5 trials\\
    A-MIL~\cite{patil2019breast} & Custom CNN & Simple & No & No & Unclear\\
    MRN~\cite{gupta2021breast} & ResNet~\cite{he2016deep} variant & Simple & No & No & Unclear\\
    MV~\cite{gupta2017breast} & Various & No & No & Voting & 5 trials\\
    SM~\cite{gupta2018sequential} & DenseNet~\cite{huang2017densely} & Simple & Pretrained & XGBoost~\cite{chen2015xgboost} & 3 trials\\
    PI~\cite{gupta2019partially} & ResNet~\cite{he2016deep} & Simple & Pretrained & Custom & 3 trials\\
    TL~\cite{deniz2018transfer} & AlexNet~\cite{krizhevsky2012imagenet} \& VGG16~\cite{simonyan2015very} & No & Pretrained & Two networks & 5-fold\\
    RPDB~\cite{man2020classification} & DenseNet~\cite{huang2017densely} & Simple & AnoGAN~\cite{schlegl2017unsupervised} & No & 5-fold CV\\
    MIM~\cite{benhammou2020breakhis} & Various & Simple & Pretrained & No & Unclear\\
    \textbf{MPCS} & Efficient-net b2~\cite{tan2019efficientnet} & Simple & Custom & No & 5-fold stratified CV\\
\hline
\end{tabular}%
}
\vspace{-6mm}
\end{table}

Another learning paradigm to effectively counter earlier stated data scarcity challenges is self-supervised learning. 
Representation learning from self-supervised learning paradigms for computer vision can be categorized majorly as (i) Joint Embedding Architecture \& Method (JEAM) (\cite{chen2020simple, grill2020bootstrap, caron2020unsupervised, zbontar2021barlow}), (ii) Prediction Methods (\cite{veeling2018rotation, noroozi2016unsupervised, doersch2015unsupervised}), and loosely (iii) Reconstruction Methods~(\cite{kingma2013auto, goodfellow2014generative}).
Specifically, JEAM can be divided further with each subdivision providing many interesting works; (i)~Contrastive Methods (PIRL~\cite{misra2020self}, SimCLR~\cite{chen2020simple}, SimCLRv2~\cite{chen2020big}, MoCo~\cite{he2020momentum}), (ii)~Distillation (BYOL~\cite{grill2020bootstrap}, SimSiam~\cite{chen2021exploring}), (iii)~Quantization (SwAV~\cite{caron2020unsupervised}, DeepCluster~\cite{caron2018deep}), and (iv)~Information Maximization (Barlow Twins~\cite{zbontar2021barlow}, VICReg~\cite{bardes2021vicreg}).
Of these divisions, this work focuses on contrastive methods.

Recently, contrastive JEAM has been tailored for medical images. Contrastive learning on digital pathology DPCL~\cite{ciga2022self} applies contrastive learning and shows improvement in the breast cancer dataset~\cite{aresta2019bach}. 
In~MICLe~\cite{azizi2021big}, which is based on SimCLR~\cite{chen2020simple}, multiple instance contrastive learning is applied by enabling input views from several image instances of the same patient.
Another work from contrastive methods on histopathology is DRL~\cite{xu2020dataefficient}.
Other applications making use of contrastive JEAM are chest X-rays~\cite{sowrirajan2021moco, liu2019align}, CT scans for COVID-19~\cite{he2020sample}, 3d-Radiomic~\cite{li2021imbalance}, and Radiograph~\cite{zhou2020comparing}.
A work using contrastive JEAM on the BreakHis dataset is SMSE~\cite{sun2021magnificationindependent}, which trains the network using pair and triplet losses.
SSL methods including JEAM require large-size data. Thus applying the contrastive JEAM paradigm on small datasets with the reduced human dependency of prior is an open challenge and interest of this work.

, 

\section{Methodology}
The primary focus of this work is to introduce a novel self-supervised pre-training method with the aim is to learn representations from data without labels, while using supervision signals from data, e.g., magnification factor and using inductive transfer from ImageNet~\cite{deng2009imagenet} pre-trained weights.


 
\subsection{Inductive Transfer Learning}



Given the fact that BreakHis~\cite{spanhol2016dataset} is a small-scale and class-imbalanced dataset, this work hypothesizes a constraint case of inductive transfer for representation learning by initializing encoder ImageNet~\cite{deng2009imagenet} pre-trained weights.
In this work, the inductive transfer (i) helps to obtain improved performance on the downstream task of malignancy classification and (ii) enables self-supervised pre-training using the proposed method on the small-scale dataset.
\subsection{Self-supervised Method - Magnification Prior Contrastive Similarity} \label{ssl}
Magnification Prior Contrastive Similarity (MPCS) method formulates self-supervised pre-training to learn representations on microscopic Histopathology WSI without labels on small-scale data. The main objective of MPCS is to lower the amount of labeled data needed for the downstream task to address challenges in supervised learning. 

MPCS construct pairs of distinct views considering characteristics of microscopic histopathology WSI (H-WSI) for contrastive similarity-based pre-training.
Microscopic H-WSI structural properties are different from natural visual macroscopic images~\cite{deng2009imagenet} (vehicles, cats, or dogs) in terms of location, size, shape, background-foreground, and concrete definition of objects.
Unlike SimCLR~\cite{chen2020simple} where pairs of distinct views from the input image is constructed by human-centered augmentations, MPCS constructs pair of distinct views using pair sampling methods based on the signal from data itself i.e. magnification factor in BreakHis~\cite{spanhol2016dataset}. 
Two H-WSI from different magnification factors of the same sample makes a pair. 
Utilizing prior from data~(magnification factor) enables meaningful contrastive learning on histopathology H-WSI and reduces dependency over human inducted prior.
Further, tumor-affected regions in H-WSI are characterized by format and highly abnormal amounts of nuclei. 
Such affected regions are promising in all the H-WSIs of different magnification for the same sample.
Thus, affected regions being common and size invariant in positive pair of a sample allow learning contrastive similarity by region attentions.


\begin{figure}[t]
    \centering
    \includegraphics[width =0.6\linewidth]{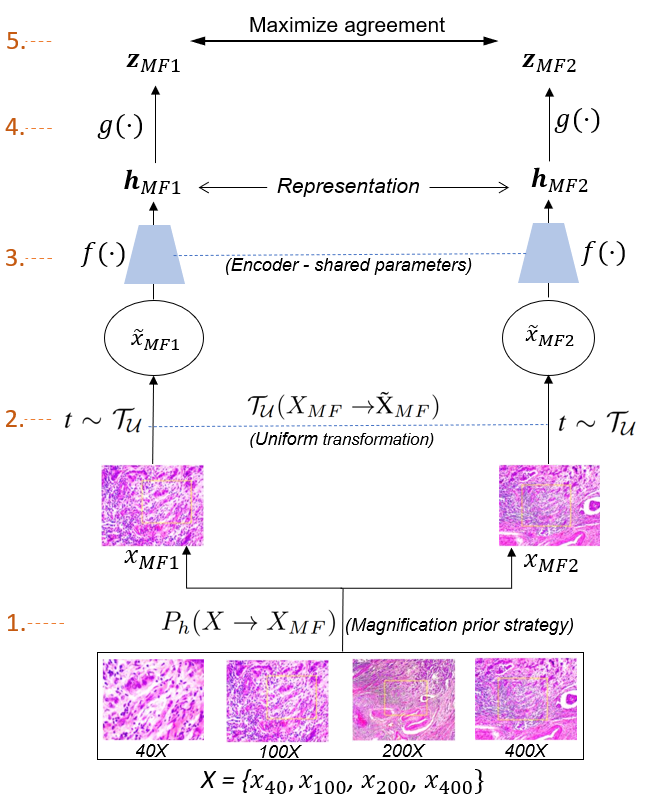}
    \vspace{-0.2cm}
    \caption{Magnification Prior Contrastive Similarity method explained}
    \label{fig:ssl_mpcs}
    \vspace{-4mm}
\end{figure}
\begin{figure}[!ht]
    \centering
    \includegraphics[width =0.6\linewidth]{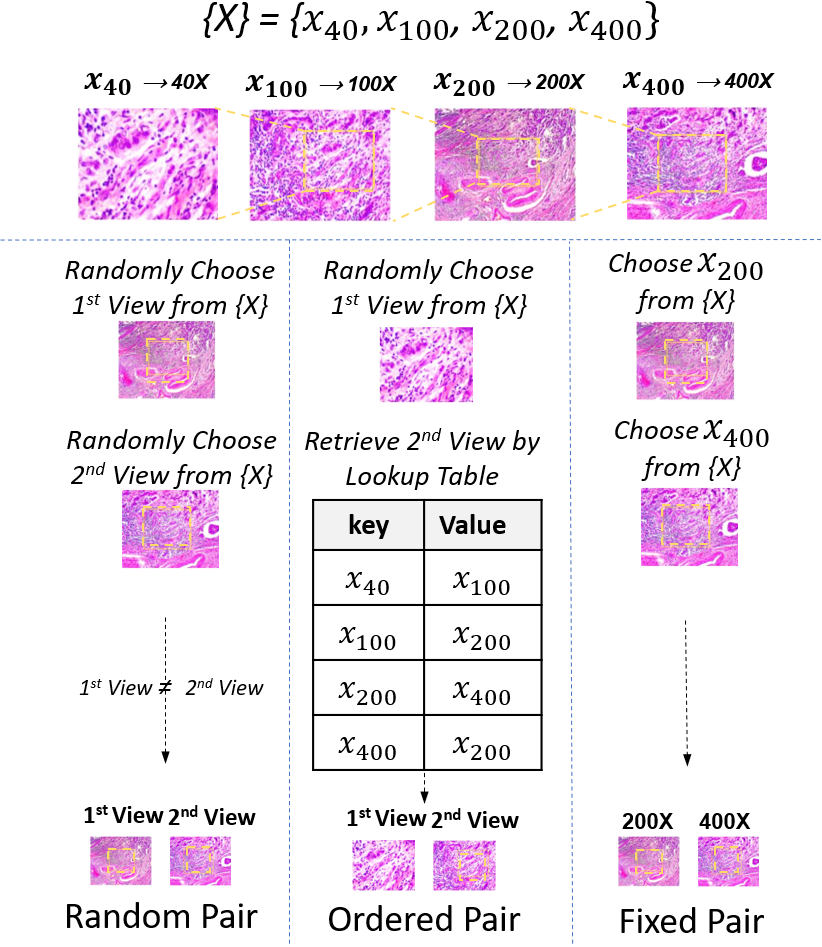}
    \vspace{-0.2cm}
    \caption{Strategies for pair sampling based on inducted Human Prior(HP). Added measure to prevent mode collapse in pre-training by $1^{st} view \neq 2^{nd} view$ in all strategies.}
    \label{fig:human_prior}
    \vspace{-5mm}
\end{figure}

\begin{figure}[!ht]
    \centering
    \includegraphics[width =0.8\linewidth]{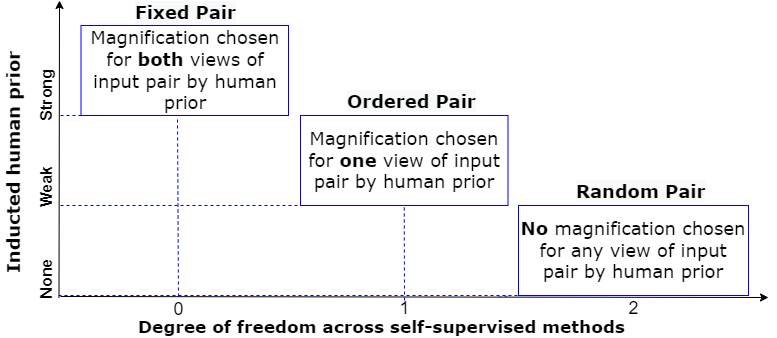}
    \vspace{-0.2cm}
    \caption{Relation between inducted Human Prior (HP) of magnification and Degree of Freedom (DoF) for method}
    \label{fig:degrre_of_freedom}
    \vspace{-4mm}
\end{figure}

Current work also hypothesizes that reduced human-prior in the pre-training method provides an enhanced degree of freedom to the method that can increase the potential of the network to learn efficient representations in a self-supervised approach. To investigate, three strategies for pair sampling are formulated based on inducted human prior. The number of human decisions defines the level of inducted human-prior (HP) during pair sampling. As explained in Figure~\ref{fig:human_prior}, in Fixed Pair, decisions to choose magnification factor for both views are by human, thus making strong human-prior. In Ordered Pair, only the second view of the pair is chosen by a human using a look-up table, making weaker human prior. In Random Pair, no human prior inducted and magnification factor for both views are sampled randomly. Further, Figure~\ref{fig:degrre_of_freedom} demonstrates the Degree of Freedom (DoF) for the method where the Fixed Pair strategy provides no DoF, Order Pair provides one DoF, and Random Pair provides 2 DoF to the method.

In MPCS, to formulate a batch of $2N$ views, randomly sampled batch of $N$ sets of input $X=\{X^{(1)}, X^{(2)}, ..., X^{(N)}\}$ are considered where each set of input $X^{(i)}=\{x^{(i)}_{40}, x^{(i)}_{100}, x^{(i)}_{200}, x^{(i)}_{400}\}$ contains the images corresponding to four magnification factors.
Positive pair of views is constructed based on a selection strategy of pair sampling which contains two views from the same example of different magnification. Further, similarity maximizes (loss minimizes) by an objective defined by the contrastive loss in Eq.~\ref{eq:loss}). MPCS is demonstrated in Figure~\ref{fig:ssl_mpcs} and components are explained below.

\begin{itemize}
    \item A \textit{domain specific human prior} module $P_h(X \to X_\mathrm{MF})$: $X=\{x_{40}, x_{100}, x_{200}, x_{400}\}$ that exploits supervision signal from data i.e. magnification and samples the two views $X_\mathrm{MF}=$(\textbf{$x_\mathrm{MF1}$}, \textbf{$x_\mathrm{MF2}$}) of different magnifications to construct pair based on employed strategy of pair sampling shown in step 1 of Figure~\ref{fig:ssl_mpcs}. \vspace{-0.2cm}
    \item A \textit{uniform stochastic transformation} based module $\displaystyle \mathcal{T_U}(X_\mathrm{MF} \to \tilde{X}_\mathrm{MF})$ that uniformly transforms both views from $X_\mathrm{MF}=$(\textbf{$x_\mathrm{MF1}$}, \textbf{$x_\mathrm{MF2}$}) to $\tilde{X}_\mathrm{MF}=$ (\textbf{$\tilde{x}_\mathrm{MF1}$}, \textbf{$\tilde{x}_\mathrm{MF2}$}) of positive pair by sampled augmentation transformations scheme shown in step 2 of Figure~\ref{fig:ssl_mpcs}.\vspace{-0.2cm}
    \item  A neural network \textit{base encoder} \textit{f}(·) which yields representations from transformed views of pair. It obtains $h_\mathrm{MF1}$ = \textit{f}($\tilde{x}_\mathrm{MF1}$) = encoder-network($\tilde{x}_\mathrm{MF1}$) and $h_\mathrm{MF2}$ = \textit{f}($\tilde{x}_\mathrm{MF2}$) = encoder-network($\tilde{x}_\mathrm{MF2}$) where $h_\mathrm{MF1}$, $h_\mathrm{MF2} \in \mathbb{R}^d$ are the output after the respective average pooling layers, shown in step 3 of Figure~\ref{fig:ssl_mpcs}.\vspace{-0.2cm}
    \item A small-scale \textit{MLP projection head} \textit{g}(·) that maps representations to the latent space where contrastive loss is applied, shown in step 4 of Figure~\ref{fig:ssl_mpcs}.
    A multi-layer perceptron with single hidden layer to obtain $z_\mathrm{MF1}$ = \textit{g}($h_\mathrm{MF1}$) = $W^{(2)}\sigma(W^{(1)}{h_\mathrm{MF1}})$ and $z_\mathrm{MF2}$ = \textit{g}($h_\mathrm{MF2}$) = $W^{(2)}\sigma(W^{(1)}{h_\mathrm{MF2}})$ where $\sigma$ is ReLU.\vspace{-0.2cm}
    \item A \textit{contrastive loss function}, normalized temperature-scaled cross entropy loss (NT-Xent) from SimCLR is defined for a contrastive prediction, shown in step 5 of Figure~\ref{fig:ssl_mpcs}.
    For given a set \textbf{$\tilde{x}_{k}$} including a positive pair examples \textbf{$\tilde{x}_\mathrm{MF1}$} and \textbf{$\tilde{x}_\mathrm{MF2}$}, the contrastive prediction task tends to find \textbf{$\tilde{x}_\mathrm{MF2}$} in \{${\mathbf{\tilde{x}_{k}}}\}_{k\neq MF1}$ for the given \textbf{$\tilde{x}_\mathrm{MF1}$}.
\end{itemize}

The loss function for a positive pair of examples (MF1, MF2) is defined as
\begin{equation}
    L_\mathrm{MF1,MF2} = -\log \dfrac{\exp(sim(\boldsymbol{z}_\mathrm{MF1},\boldsymbol{z}_\mathrm{MF2})/\tau)} {\sum_{k =1}^{2N} 1_{[k \neq MF1]} exp(sim(\boldsymbol{z}_\mathrm{MF1},\boldsymbol{z}_{k})/\tau)}
    \label{eq:loss}
\end{equation}

In Eq.~(\ref{eq:loss}), where $1_{[k \neq MF1]} \in {0, 1}$ is an indicator evaluating to 1 if $k \neq i$.
  \label{methdology}


\section{Experimental Evaluations}
This section investigates the representation learning capabilities of the proposed method MPCS on two encoder networks with detailed experimentation on three public datasets.
\subsection{Datasets}
\subsubsection{BreakHis}
BreakHis~\cite{spanhol2016dataset} dataset consists of 2,480 benign and 5,429 malignant histopathological microscopic images from 82 patients at four magnification levels (40×, 100×, 200×, 400×). Each image in the BreakHis dataset is of size 700×460, stained with hematoxylin and eosin (HE). Following the previous works, two evaluation metrics are used, image-level accuracy(ILA) and patient-level accuracy (PLA). PLA shows patient-wise classification performance, calculated as the mean over total no. of patients using patient-score. Patient-score is correctly classified images of the patient over a total number of images of that patient. ILA disregards patient-level details and thus serves as standard image classification accuracy.    
\subsubsection{BACH}
The second dataset, Breast Cancer Histology Images (BACH)~\cite{aresta2019bach} is from the ICIAR2018 Grand Challenge and contains 400 histopathology slides. The BACH dataset has four classes, normal, benign, in-situ, and invasive. The slide size is relatively large, 2048 × 1536 pixels; thus, patches of size 512x512. Two evaluation metrics, patch-wise accuracy and image-wise accuracy, are used, whereas image-wise accuracy is calculated based on majority voting over the patches of the respective image.
\subsubsection{Breast Cancer Cell Dataset}
The third dataset, Breast Cancer Cell Dataset~\cite{gelasca2008evaluation} is from the University of California, Santa Barbara Biosegmentation Benchmark. This dataset contains 58 HE-stained histopathology 896x768 size images of breast tissue, of which 26 are malignant, and 32 are benign. Patches of size 224x224 were created, and image-wise accuracy was calculated using majority voting over patches of the respective image. 
\subsection{Encoder Architectures}
In this current work, the proposed method MPCS investigated two different CNN encoder architectures. ResNet-50~\cite{yu2017dilated} and Efficient-net b2~\cite{tan2019efficientnet} are used for pre-training and fine-tuning. SSL-specific MLP projection head used for Efficient-net b2 is three layers network of 2048-1204-128 units, whereas ResNet-50 is the most common backbone encoder, used projection head adapted from SimCLR, having 1024-128 units. encoder and projection heads are demonstrated in Fig.~\ref{fig:ssl_mpcs}.
\subsection{Training Protocol}
This section shares parameter configurations used in pre-training and fine-tuning. 
\subsubsection{SSL pre-training}
Self-supervised pre-trainings of both encoders take place on the BreakHis dataset for 1000 epochs with temperature parameter $0.01$, learning rate 1e-05, and a set of augmentation methods such as color-jitter, flip, and rotation. Efficient-net b2 encoder pre-trained using Adam optimizer with a batch size of 128 and image input of $(341, 341)$. However, ResNet-50 adapted standard configurations of self-supervised practices and pre-trained using the LARS optimizer with a batch size of 1024 and input image size of 224x224.
\subsubsection{Fine-tuning}
The common training configurations across datasets for both encoders are learning rate of 2e-05, batch size of 32, image input of 224x224, augmentations methods such as random crop, flip, affine, rotation, and color-jitter, and using adam as optimizer. A dropout of 0.3 is used in the fully connected layer.
\subsection{Experimentation Details}

To ensure the reliability and consistency of the models, this work follows 5-cross validation data split scheme. This is applied to all three datasets in which each fold contained 20\% data, following class distribution from whole data. Four out of five folds are used for training \& validation, and the remaining one for testing. Thus all the results reported are in terms of mean value with standard deviation.
In the above-stated 5-cross validation settings, both backbone encoders, ResNet-50 and Efficient-net b2, are being pre-trained on the first dataset BreakHis with all three variants (ordered pair, random pair, and fixed pair) of the proposed SSL method MPCS for learning domain-specific representations. Further, downstream-task-specific fine-tuning experiments are carried out to investigate the impact of learned representations for all three datasets e.g., BreakHis, BACH, and Breast Cancer Cell. 
Following are the details of experimentation for each dataset.
Table~\ref{tab:experiments_details} describes fine-tuning experiments for the first dataset BreakHis. All the mentioned experiments for malignancy classification are conducted for both encoders, Efficient-net b2 and ResNet-50 for all four magnifications (40X, 100X, 200X, and 400X). Experiments (Exp-1 to Exp-4) in \textit{Limited Labeled Data Setting} evaluate model performance while using only 20\% labels, whereas experiments (Exp-5 to Exp-8) in \textit{Fully Supervised Setting} use all labels. The sole objective is to compare the performance of models pre-trained on proposed MPCS methods against ImageNet~\cite{deng2009imagenet} pre-trained model to analyze the effect of learned representations. Preferred names in discussion used as MPCS-X for ImageNet $\to$ MPCS-X.
Experiments(Exp-9 to Exp-12) for the second dataset BACH~\cite{aresta2019bach} are described in Table~\ref{tab:exp_bach} based on the ResNet-50 encoder. All the images are divided into 512X512 size patches; thus, performance is measured patch-wise and image-wise (using majority voting suggested in~\cite{vesal2018classification}). The major objectives are 1) evaluating pre-trained models from the proposed method against the ImageNet-based transfer learning approach~\cite{aresta2019bach} 2) evaluating the ability to learn downstream tasks with limited labels ranging from 5\% to 100\% of labels from the train data portion.
Finally, to evaluate the effect of learned domain-specific representations on small-scale data, a series of fine-tuning (all layers trained) and linear evaluation experiments (only fully-connected layers trainable) Exp-13 to Exp-16 are conducted Breast Cancer Cell dataset and described in Table~\ref{tab:exp_bisque}. Similar to BACH dataset, images for this dataset were also divided into sizes 224X224 for training and test, and performance was measured likewise.
\begin{table}[t]
\caption{Experiment details for BreakHis dataset}
\vspace{-0.3cm}
\label{tab:experiments_details}
\resizebox{\columnwidth}{!}{%
\begin{tabular}{ccccc}
\hline
\multicolumn{1}{c|}{\multirow{2}{*}{No.}} & \multicolumn{1}{c|}{\multirow{2}{*}{Pre-training Method}}          & \multicolumn{1}{c|}{\multirow{2}{*}{\begin{tabular}[c]{@{}c@{}}BreakHis data\\  (\%)for SSL\end{tabular}}} & \multicolumn{2}{c}{\begin{tabular}[c]{@{}c@{}}Finetuning on \\ BreakHis Dataset\end{tabular}} \\ \cline{4-5} 
\multicolumn{1}{c|}{}                       & \multicolumn{1}{c|}{}                                              & \multicolumn{1}{c|}{}                                                                                      & \multicolumn{1}{c|}{Train (\%)}                          & Test (\%)                          \\ \hline
\multicolumn{5}{c}{Limited Labeled Data Setting}                                                                                                                                                                                                                                                                             \\ \hline
\multicolumn{1}{c|}{Exp-1}                      & \multicolumn{1}{c|}{ImageNet}                                      & \multicolumn{1}{c|}{-}                                                                                     & \multicolumn{1}{c|}{20\%}                                & 20\%                               \\
\multicolumn{1}{c|}{Exp-2}                      & \multicolumn{1}{c|}{ImageNet $\to$ MPCS-Fixed Pair}   & \multicolumn{1}{c|}{60\%}                                                                                  & \multicolumn{1}{c|}{20\%}                                & 20\%                               \\
\multicolumn{1}{c|}{Exp-3}                      & \multicolumn{1}{c|}{ImageNet $\to$ MPCS-Ordered Pair} & \multicolumn{1}{c|}{60\%}                                                                                  & \multicolumn{1}{c|}{20\%}                                & 20\%                               \\
\multicolumn{1}{c|}{Exp-4}                      & \multicolumn{1}{c|}{ImageNet $\to$ MPCS-Random Pair}  & \multicolumn{1}{c|}{60\%}                                                                                  & \multicolumn{1}{c|}{20\%}                                & 20\%                               \\ \hline
\multicolumn{5}{c}{Fully Supervised Data Setting}                                                                                                                                                                                                                                                                             \\ \hline
\multicolumn{1}{c|}{Exp-5}                      & \multicolumn{1}{c|}{ImageNet}                                      & \multicolumn{1}{c|}{-}                                                                                     & \multicolumn{1}{c|}{80\%}                                & 20\%                               \\
\multicolumn{1}{c|}{Exp-6}                      & \multicolumn{1}{c|}{ImageNet $\to$ MPCS-Fixed Pair} & \multicolumn{1}{c|}{60\%}                                                                                  & \multicolumn{1}{c|}{80\%}                                & 20\%                               \\
\multicolumn{1}{c|}{Exp-7}                      & \multicolumn{1}{c|}{ImageNet $\to$ MPCS-Ordered Pair}  & \multicolumn{1}{c|}{60\%}                                                                                  & \multicolumn{1}{c|}{80\%}                                & 20\%                               \\
\multicolumn{1}{c|}{Exp-8}                      & \multicolumn{1}{c|}{ImageNet $\to$ MPCS-Random Pair}  & \multicolumn{1}{c|}{60\%}                                                                                  & \multicolumn{1}{c|}{80\%}                                & 20\%                               \\ \hline
\end{tabular}%
}
\vspace{-2mm}
\end{table}

\begin{table}[t]
\caption{Experiment details for BACH dataset}
\label{tab:exp_bach}
\resizebox{\columnwidth}{!}{%
\begin{tabular}{c|c|cc}
\hline
                      &                                                                                          & \multicolumn{2}{c}{Fine-tuning on BACH dataset}                                                                   \\ \cline{3-4} 
\multirow{-2}{*}{No.} & \multirow{-2}{*}{\begin{tabular}[c]{@{}c@{}}Pre-training \\ method/weights\end{tabular}} & \multicolumn{1}{c|}{Labels(\%) from Train Data}                                                 & Test data (\%)              \\ \hline
Exp-9                 & ImageNet~\cite{vesal2018classification}-re-implement                                                                                 & \multicolumn{1}{c|}{{\color[HTML]{1E1E1E} 100\% (~80\% train data)}}                                    & {\color[HTML]{1E1E1E} 20\%} \\
Exp-10                & MPCS-Fixed Pair (BreakHis)                                                               & \multicolumn{1}{c|}{{\color[HTML]{1E1E1E} {[}5\%, 10\%, 20\%, 40\%, 60\%, 80\%, 100\%{]}}} & {\color[HTML]{1E1E1E} 20\%} \\
Exp-11                & MPCS-Ordered Pair(BreakHis)                                                              & \multicolumn{1}{c|}{{\color[HTML]{1E1E1E} {[}5\%, 10\%, 20\%, 40\%, 60\%, 80\%, 100\%{]}}} & 20\%                        \\
Exp-12                & MPCS-Random Pair(BreakHis)                                                               & \multicolumn{1}{c|}{{\color[HTML]{1E1E1E} {[}5\%, 10\%, 20\%, 40\%, 60\%, 80\%, 100\%{]}}} & {\color[HTML]{1E1E1E} 20\%} \\ \hline
\end{tabular}%
}
\vspace{-3mm}
\end{table}

\begin{table}[t]
\caption{Experiment details for Breast Cancer Cell Dataset}
\label{tab:exp_bisque}
\resizebox{\columnwidth}{!}{%
\begin{tabular}{c|c|cc|cc}
\hline
                      &                                                                                          & \multicolumn{2}{c|}{\begin{tabular}[c]{@{}c@{}}Fine-tuning on \\ Breast Cancer Cell Dataset\end{tabular}} & \multicolumn{2}{c}{\begin{tabular}[c]{@{}c@{}}Linear-evaluation on  \\ Breast Cancer Cell Dataset\end{tabular}} \\ \cline{3-6} 
\multirow{-2}{*}{No.} & \multirow{-2}{*}{\begin{tabular}[c]{@{}c@{}}Pre-training \\ method/weights\end{tabular}} & \multicolumn{1}{c|}{Train data(\%)}                            & Test data (\%)                           & \multicolumn{1}{c|}{Train data(\%)}                            & \multicolumn{1}{l}{Test data(\%)}              \\ \hline
Exp-13                & MPCS-Fixed Pair (BreakHis)                                                               & \multicolumn{1}{c|}{{\color[HTML]{1E1E1E} 80\%}}               & {\color[HTML]{1E1E1E} 20\%}              & \multicolumn{1}{c|}{{\color[HTML]{1E1E1E} 80\%}}               & {\color[HTML]{1E1E1E} 20\%}                    \\
Exp-14                & MPCS-Ordered Pair(BreakHis)                                                              & \multicolumn{1}{c|}{{\color[HTML]{1E1E1E} 80\%}}               & 20\%                                     & \multicolumn{1}{c|}{{\color[HTML]{1E1E1E} 80\%}}               & 20\%                                           \\
Exp-15                & MPCS-Random Pair(BreakHis)                                                               & \multicolumn{1}{c|}{{\color[HTML]{1E1E1E} 80\%}}               & {\color[HTML]{1E1E1E} 20\%}              & \multicolumn{1}{c|}{{\color[HTML]{1E1E1E} 80\%}}               & {\color[HTML]{1E1E1E} 20\%}                    \\ \hline
\end{tabular}%
}
\vspace{-3mm}
\end{table}

\section{Results \& Discussions}

\begin{table*}[t]
\centering
\caption{Performance evaluation of the proposed methods in limited labelled data setting when fine-tuning only on 20\% labeled data. (across magnification, p \textless 0.05 for all three self-supervised methods for PLA). Ablation results for 40\%, 60\%, and 80\% labels are provided in supplementary content.}
\label{ft_20}
\begin{scriptsize}
\begin{tabular}{c|c|c|c|c|c|c|c}
\hline
Encoder                            & Metric                                                                              & Method       & 40X                                        & 100X                                       & 200X                                       & 400X                                       & Mean                                                   \\ \hline
                                   &                                                                                     & ImageNet     & {\color[HTML]{1E1E1E} 86.36±6.13}          & {\color[HTML]{1E1E1E} 87.80±4.15}          & {\color[HTML]{1E1E1E} 86.82±5.74}          & {\color[HTML]{1E1E1E} 84.98±6.05}          & 86.49±5.63                                             \\
                                   &                                                                                     & FixedPair    & {\color[HTML]{1E1E1E} 87.26±4.46}          & {\color[HTML]{1E1E1E} 87.45±2.35}          & {\color[HTML]{1E1E1E} 89.38±2.18}          & {\color[HTML]{1E1E1E} 88.12±3.84}          & \multicolumn{1}{l}{{\color[HTML]{444444} 88.05 ±3.20}} \\
                                   &                                                                                     & Ordered Pair & {\color[HTML]{1E1E1E} \textbf{87.40±3.73}} & {\color[HTML]{1E1E1E} \textbf{89.30±2.74}} & {\color[HTML]{1E1E1E} \textbf{90.50±2.19}} & {\color[HTML]{1E1E1E} \textbf{88.35±3.30}} & \textbf{88.89±2.99}                                    \\
                                   & \multirow{-4}{*}{\begin{tabular}[c]{@{}c@{}}Image Level \\ Accuracy\end{tabular}}   & Random Pair  & {\color[HTML]{1E1E1E} 86.21±4.20}          & {\color[HTML]{1E1E1E} 89.55±2.84}          & {\color[HTML]{1E1E1E} 89.18±4.05}          & {\color[HTML]{1E1E1E} 87.34±3.44}          & 88.07±3.63                                             \\ \cline{2-8} 
                                   &                                                                                     & ImageNet     & {\color[HTML]{1E1E1E} 86.13±5.15}          & {\color[HTML]{1E1E1E} 87.76±6.03}          & {\color[HTML]{1E1E1E} 85.79±4.10}          & {\color[HTML]{1E1E1E} 85.51±5.27}          & 86.30±5.14                                             \\
                                   &                                                                                     & FixedPair    & {\color[HTML]{1E1E1E} 86.90±4.14}          & {\color[HTML]{1E1E1E} 87.64±3.05}          & {\color[HTML]{1E1E1E} 89.60±3.28}          & {\color[HTML]{1E1E1E} 88.26±3.05}          & \multicolumn{1}{l}{{\color[HTML]{444444} 88.10 ±3.38}} \\
                                   &                                                                                     & Ordered Pair & {\color[HTML]{1E1E1E} \textbf{87.19±3.20}} & {\color[HTML]{1E1E1E} \textbf{88.86±2.58}} & {\color[HTML]{1E1E1E} \textbf{90.20±3.26}} & {\color[HTML]{1E1E1E} \textbf{88.96±3.22}} & \textbf{88.80±3.07}                                    \\
\multirow{-8}{*}{Efficient-net b2} & \multirow{-4}{*}{\begin{tabular}[c]{@{}c@{}}Patient Level\\  Accuracy\end{tabular}} & Random Pair  & {\color[HTML]{1E1E1E} 87.17±3.88}          & {\color[HTML]{1E1E1E} 88.36±2.84}          & {\color[HTML]{1E1E1E} 88.58±4.01}          & {\color[HTML]{1E1E1E} 88.66±3.13}          & 88.19±3.48                                             \\ \hline
                                   &                                                                                     & ImageNet     & {\color[HTML]{1E1E1E} 87.40±4.88}          & {\color[HTML]{1E1E1E} 86.22±5.71}          & {\color[HTML]{1E1E1E} 86.02±4.74}          & {\color[HTML]{1E1E1E} 85.30±5.95}          & 86.24±5.32                                             \\
                                   &                                                                                     & FixedPair    & {\color[HTML]{1E1E1E} 86.69±3.96}          & {\color[HTML]{1E1E1E} 86.94±3.05}          & {\color[HTML]{1E1E1E} 88.76±2.28}          & {\color[HTML]{1E1E1E} \textbf{88.81±2.77}} & \multicolumn{1}{l}{{\color[HTML]{444444} 87.68 ±3.01}} \\
                                   &                                                                                     & Ordered Pair & {\color[HTML]{1E1E1E} \textbf{87.56±3.48}} & {\color[HTML]{1E1E1E} \textbf{88.60±3.01}} & {\color[HTML]{1E1E1E} \textbf{89.77±2.19}} & {\color[HTML]{1E1E1E} 87.61±3.48}          & \textbf{88.38±3.04}                                    \\
                                   & \multirow{-4}{*}{\begin{tabular}[c]{@{}c@{}}Image Level \\ Accuracy\end{tabular}}   & Random Pair  & {\color[HTML]{1E1E1E} 87.06±3.40}          & {\color[HTML]{1E1E1E} 87.96±3.44}          & {\color[HTML]{1E1E1E} 88.55±3.15}          & {\color[HTML]{1E1E1E} 86.64±2.98}          & 87.55±3.24                                             \\ \cline{2-8} 
                                   &                                                                                     & ImageNet     & {\color[HTML]{1E1E1E} 87.10±4.80}          & {\color[HTML]{1E1E1E} 88.06±5.11}          & {\color[HTML]{1E1E1E} 84.19±4.28}          & {\color[HTML]{1E1E1E} 85.01±5.27}          & 86.09±4.86                                             \\
                                   &                                                                                     & FixedPair    & {\color[HTML]{1E1E1E} 87.45±3.96}          & {\color[HTML]{1E1E1E} 86.38±3.12}          & {\color[HTML]{1E1E1E} 88.18±3.00}          & {\color[HTML]{1E1E1E} \textbf{88.89±2.98}} & \multicolumn{1}{l}{{\color[HTML]{444444} 87.72 ±3.27}} \\
                                   &                                                                                     & Ordered Pair & {\color[HTML]{1E1E1E} \textbf{87.88±2.89}} & {\color[HTML]{1E1E1E} \textbf{88.21±3.21}} & {\color[HTML]{1E1E1E} \textbf{89.52±3.26}} & {\color[HTML]{1E1E1E} 87.90±3.03}          & \textbf{88.38±3.09}                                    \\
\multirow{-8}{*}{ResNet-50}        & \multirow{-4}{*}{\begin{tabular}[c]{@{}c@{}}Patient Level\\  Accuracy\end{tabular}} & Random Pair  & {\color[HTML]{1E1E1E} 87.17±2.98}          & {\color[HTML]{1E1E1E} 87.96±3.02}          & {\color[HTML]{1E1E1E} 88.76±3.55}          & {\color[HTML]{1E1E1E} 88.06±3.00}          & 87.99±3.14                         \\ \hline                    
\end{tabular}
\end{scriptsize}
\end{table*} 


\begin{table*}[ht]
\caption{Comparison of proposed methods with state-of-the-art methods on downstream task of classifying histopathological images at different magnification factors in fully supervised setting(using 100\% train set labels). RN-50 indicates ResNet-50 and Eff-net b2 indicates Efficient-net b2 encoder. Also, OP indicates Ordered Pair, FP indicates Fixed Pair, and RP indcates Random Pair.}
\label{ft_80_sota}
\begin{adjustbox}{max width=\textwidth}
\begin{tabular}{c|cccc|c|cccc|c}

\hline
                                                            & \multicolumn{4}{c|}{Patient Level Accuracy (RR)}                                                                                                                                                                           &                        & \multicolumn{4}{c|}{Image Level Accuracy}                                                                                                            &                                                       \\ \cline{2-5} \cline{7-10}
\multirow{-2}{*}{Method}                                    & \multicolumn{1}{c|}{40X}                 & \multicolumn{1}{c|}{100X}                                        & \multicolumn{1}{c|}{200X}                                       & 400X                                       & \multirow{-2}{*}{Mean} & \multicolumn{1}{c|}{40X}                 & \multicolumn{1}{c|}{100X}                & \multicolumn{1}{c|}{200X}                & 400X                & \multirow{-2}{*}{Mean}                                \\
\hline
Original-GLCM\cite{spanhol2016breast}   & \multicolumn{1}{c|}{74.7±1.0}            & \multicolumn{1}{c|}{78.6±2.6}                                    & \multicolumn{1}{c|}{83.4±3.3}                                   & 81.7±3.3                                   & 79.60±2.55             & \multicolumn{1}{c|}{-}                   & \multicolumn{1}{c|}{-}                   & \multicolumn{1}{c|}{-}                   & -                   & -                                                     \\
PFTAS\cite{hamilton2007fast}               & \multicolumn{1}{c|}{83.80±2.0}           & \multicolumn{1}{c|}{82.10±4.9}                                   & \multicolumn{1}{c|}{85.10±3.1}                                  & 82.30±3.8                                  & 83.33±3.45             & \multicolumn{1}{c|}{-}                   & \multicolumn{1}{c|}{-}                   & \multicolumn{1}{c|}{-}                   & -                   & -                                                     \\
MIL-NP\cite{sudharshan2019multiple}        & \multicolumn{1}{c|}{92.1±5.9}            & \multicolumn{1}{c|}{89.1±5.2}                                    & \multicolumn{1}{c|}{87.2±4.3}                                   & 82.7±3.0                                   & 87.77±4.6              & \multicolumn{1}{c|}{87.8±5.6}            & \multicolumn{1}{c|}{85.6±4.3}            & \multicolumn{1}{c|}{80.8±2.8}            & 82.9±4.1            & \multicolumn{1}{l}{{\color[HTML]{444444} 84.28±4.20}} \\
SW\cite{spanhol2016breast}                 & \multicolumn{1}{c|}{88.6±5.6}            & \multicolumn{1}{c|}{84.5±2.4}                                    & \multicolumn{1}{c|}{85.3±3.8}                                   & 81.7±4.9                                   & 85.02±4.17             & \multicolumn{1}{c|}{89.6±6.58}           & \multicolumn{1}{c|}{85.0±4.8}            & \multicolumn{1}{c|}{84.0±3.2}            & 80.8±3.1            & \multicolumn{1}{l}{{\color[HTML]{444444} 84.85±4.42}} \\
MI\cite{bayramoglu2016deep}                & \multicolumn{1}{c|}{83.08±2.08}          & \multicolumn{1}{c|}{83.17±3.51}                                  & \multicolumn{1}{c|}{84.63±2.72}                                 & 82.10±4.42                                 & 83.25±3.18             & \multicolumn{1}{c|}{-}                   & \multicolumn{1}{c|}{-}                   & \multicolumn{1}{c|}{-}                   & -                   & -                                                     \\
Deep\cite{spanhol2017deep}                 & \multicolumn{1}{c|}{84.0±6.9}            & \multicolumn{1}{c|}{83.9±5.9}                                    & \multicolumn{1}{c|}{86.3±3.5}                                   & 82.1±2.4                                   & 84.07±4.67             & \multicolumn{1}{c|}{84.6±2.9}            & \multicolumn{1}{c|}{84.8±4.2}            & \multicolumn{1}{c|}{84.2±1.7}            & 81.6±3.7            & \multicolumn{1}{l}{{\color[HTML]{444444} 83.80±3.13}} \\
MILCNN\cite{sudharshan2019multiple}        & \multicolumn{1}{c|}{86.9±5.4}            & \multicolumn{1}{c|}{85.7±4.8}                                    & \multicolumn{1}{c|}{85.9±3.9}                                   & 83.4±5.3                                   & 85.47±4.85             & \multicolumn{1}{c|}{86.1 ± 4.28}         & \multicolumn{1}{c|}{83.8±3.0}            & \multicolumn{1}{c|}{80.2±2.6}            & 80.6±4.6            & \multicolumn{1}{l}{{\color[HTML]{444444} 82.68±3.62}} \\
GLPB\cite{ataky2020data}                   & \multicolumn{1}{c|}{84.5±4.2}            & \multicolumn{1}{c|}{83.5±2.0}                                    & \multicolumn{1}{c|}{89.6±5.0}                                   & \textbf{88.2±4.0}                          & 86.45±3.8              & \multicolumn{1}{c|}{82.1±6.4}            & \multicolumn{1}{c|}{81.4±4.8}            & \multicolumn{1}{c|}{88.4±5.0}            & 87.2±4.5            & \multicolumn{1}{l}{{\color[HTML]{444444} 84.78±5.18}} \\
RPDB\cite{man2020classification}$^\Cross$  & \multicolumn{1}{c|}{92.02±0.9}           & \multicolumn{1}{c|}{90.21±2.40}                                  & \multicolumn{1}{c|}{81.94±1.70}                                 & 80.09±0.70                                 & 88.06±1.4              & \multicolumn{1}{c|}{\textbf{94.26±3.2}}  & \multicolumn{1}{c|}{92.71±0.4}           & \multicolumn{1}{c|}{83.90±2.8}           & 82.74±1.5           & \multicolumn{1}{l}{{\color[HTML]{444444} 88.40±1.98}} \\
SMSE\cite{sun2021magnificationindependent} & \multicolumn{1}{c|}{87.51±4.07}          & \multicolumn{1}{c|}{89.12±2.86}                                  & \multicolumn{1}{c|}{90.83±3.31}                                 & 87.10±3.80                                 & 88.64±3.51             & \multicolumn{1}{c|}{-}                   & \multicolumn{1}{c|}{-}                   & \multicolumn{1}{c|}{-}                   & -                   & -                                                     \\ \hline
ImageNet (Eff-net b2)                                       & \multicolumn{1}{c|}{91.91±4.25}          & \multicolumn{1}{c|}{91.93±4.20}                                  & \multicolumn{1}{c|}{91.46±5.17}                                 & 88.10±3.88                                 & 90.85±4.36             & \multicolumn{1}{c|}{92.12±4.18}          & \multicolumn{1}{c|}{92.66±4.20}          & \multicolumn{1}{c|}{91.83±4.55}          & 88.35±5.21          & 91.24±4.54                                            \\
MPCS-FP (Eff-net b2)                                        & \multicolumn{1}{c|}{92.23±3.50}          & \multicolumn{1}{c|}{92.72±3.68}                                  & \multicolumn{1}{c|}{91.94±3.80}                                 & 88.40±3.26                                 & 91.33±3.56             & \multicolumn{1}{c|}{92.23±3.80}          & \multicolumn{1}{c|}{\textbf{93.57±3.23}} & \multicolumn{1}{c|}{92.23±2.98}          & 88.40±3.90          & 91.61±3.48                                            \\
MPCS-OP (Eff-net b2)                                        & \multicolumn{1}{c|}{\textbf{92.45±3.25}} & \multicolumn{1}{c|}{\textbf{93.47±2.98}}                         & \multicolumn{1}{c|}{{\color[HTML]{1E1E1E} \textbf{92.44±3.30}}} & {\color[HTML]{1E1E1E} 89.00±3.05}          & \textbf{91.84±3.15}    & \multicolumn{1}{c|}{92.67±3.36}          & \multicolumn{1}{c|}{\textbf{93.63±3.38}} & \multicolumn{1}{c|}{\textbf{92.72±2.80}} & \textbf{88.74±3.90} & \textbf{91.94±3.36}                                   \\
MPCS-RP (Eff-net b2)                                        & \multicolumn{1}{c|}{\textbf{93.26±3.48}} & \multicolumn{1}{c|}{{\color[HTML]{1E1E1E} \textbf{93.57± 3.36}}} & \multicolumn{1}{c|}{{\color[HTML]{1E1E1E} \textbf{92.23±3.21}}} & {\color[HTML]{1E1E1E} \textbf{89.57±3.79}} & \textbf{92.15±3.46}    & \multicolumn{1}{c|}{\textbf{93.45±3.55}} & \multicolumn{1}{c|}{93.38±2.80}          & \multicolumn{1}{c|}{\textbf{92.28±3.49}} & \textbf{89.81±3.15} & \textbf{92.23±3.24}                                   \\
ImageNet (RN-50)                                            & \multicolumn{1}{c|}{91.46±4.30}          & \multicolumn{1}{c|}{91.24±5.1}                                   & \multicolumn{1}{c|}{90.72±4.68}                                 & 87.90±4.12                                 & 90.33±4.55             & \multicolumn{1}{c|}{91.83±5.12}          & \multicolumn{1}{c|}{92.23±4.15}          & \multicolumn{1}{c|}{91.61±4.00}          & 87.88±4.80          & 90.89±4.52                                            \\
MPCS-FP (RN-50)                                             & \multicolumn{1}{c|}{91.83±3.88}          & \multicolumn{1}{c|}{92.67±2.72}                                  & \multicolumn{1}{c|}{91.61±3.40}                                 & 89.00±3.15                                 & 91.28±3.29             & \multicolumn{1}{c|}{92.24±3.48}          & \multicolumn{1}{c|}{92.66±3.88}          & \multicolumn{1}{c|}{91.91±3.68}          & 88.40±3.66          & 91.30±3.68                                            \\
MPCS-OP (RN-50)                                             & \multicolumn{1}{c|}{\textbf{93.00±3.66}} & \multicolumn{1}{c|}{93.26±3.08}                                  & \multicolumn{1}{c|}{{\color[HTML]{1E1E1E} \textbf{92.28±2.88}}} & {\color[HTML]{1E1E1E} \textbf{88.74±3.60}} & \textbf{91.82±3.31}    & \multicolumn{1}{c|}{\textbf{93.26±3.40}} & \multicolumn{1}{c|}{\textbf{93.45±2.89}} & \multicolumn{1}{c|}{\textbf{92.45±3.77}} & \textbf{89.57±2.96} & \textbf{92.18±3.26}                                   \\
MPCS-RP (RN-50)                                             & \multicolumn{1}{c|}{92.72±3.50}          & \multicolumn{1}{c|}{{\color[HTML]{1E1E1E} \textbf{93.57± 2.88}}} & \multicolumn{1}{c|}{{\color[HTML]{1E1E1E} 92.23±3.90}}          & {\color[HTML]{1E1E1E} 88.40±3.05}          & 91.73±3.33             & \multicolumn{1}{c|}{92.72±3.38}          & \multicolumn{1}{c|}{92.72±4.02}          & \multicolumn{1}{c|}{91.91±3.21}          & 88.56±3.89          & 91.48±3.66                                            \\ \hline
\end{tabular}
\end{adjustbox}
\end{table*}
 
\begin{table}[t]
\caption{Performance comparison of proposed method MPCS (ResNet-50 encoder) on BACH dataset with other state-of-the-arts for four class classification. RN-50 indicates ResNet-50.}
\resizebox{\columnwidth}{!}{%
\label{tab:bach_results_sota_compare}
\begin{tabular}{c|cc|cc}
\hline
                                                                & \multicolumn{2}{c|}{Image-wise accuracy}                              & \multicolumn{2}{c}{Patch-wise accuracy}                               \\ \cline{2-5} 
\multirow{-2}{*}{Method}                                        & validation                        & test                              & validation                        & test                              \\ \hline
PT~\cite{roy2019patch}                                                              & {\color[HTML]{1E1E1E} -}          & {\color[HTML]{1E1E1E} 90.00}      & {\color[HTML]{1E1E1E} -}          & {\color[HTML]{1E1E1E} 77.40}      \\
HN~\cite{yan2020breast} (RN-50)                                                      & {\color[HTML]{1E1E1E} -}          & {\color[HTML]{1E1E1E} 81.60}      & {\color[HTML]{1E1E1E} -}          & {\color[HTML]{1E1E1E} -}          \\
HN~\cite{yan2020breast}                                                              & {\color[HTML]{1E1E1E} -}          & {\color[HTML]{1E1E1E} 91.30}      & {\color[HTML]{1E1E1E} -}          & {\color[HTML]{1E1E1E} 82.10}      \\
DPCL~\cite{ciga2022self}                                                            & -                                 & {\color[HTML]{1E1E1E} 87.00}      & -                                 & -                                 \\
\begin{tabular}[c]{@{}c@{}}ImageNet~\cite{vesal2018classification}\\ re-implement\end{tabular} & {\color[HTML]{1E1E1E} 92.40±2.04} & {\color[HTML]{1E1E1E} 90.50±2.10} & {\color[HTML]{1E1E1E} 80.56±3.06} & {\color[HTML]{1E1E1E} 80.00±2.64} \\
\hline
MPCS-FP                                                         & {\color[HTML]{1E1E1E} 92.50±1.90} & {\color[HTML]{1E1E1E} 90.55±2.05} & {\color[HTML]{1E1E1E} 84.25±1.88} & {\color[HTML]{1E1E1E} 82.79±2.05} \\
MPCS-OP                                                         & {\color[HTML]{1E1E1E} \textbf{93.31±1.85}} & {\color[HTML]{1E1E1E} \textbf{91.85±1.77}} & {\color[HTML]{1E1E1E} \textbf{83.90±1.89}} & {\color[HTML]{1E1E1E} \textbf{83.13±2.00}} \\
MPCS-RP                                                         & {\color[HTML]{1E1E1E} 93.00±1.88} & {\color[HTML]{1E1E1E} 91.00±2.32} & {\color[HTML]{1E1E1E} 83.78±2.09} & {\color[HTML]{1E1E1E} 82.90±2.10} \\ \hline
\end{tabular}%
}
\vspace{-6mm}
\end{table}
\begin{table}[t]
\caption{Performance evaluation (image-wise accuracy and f1 score) of proposed method MPCS on BACH dataset for limited label range for ResNet-50 encoder.}
\label{tab:bach_results_labelwise}
\resizebox{\columnwidth}{!}{%
\begin{tabular}{c|ccc|ccc}
\hline
\begin{tabular}[c]{@{}c@{}}Label(\%) from\\ train data\end{tabular} & \multicolumn{3}{c|}{F1 score (test data)}                                                              & \multicolumn{3}{c}{Accuracy (test data)}                                                                  \\ \hline
                                                                    & MPCS-FP                          & MPCS-OP                          & MPCS-RP                          & MPCS-FP                           & MPCS-OP                           & MPCS-RP                           \\ \hline
5\%                                                                 & {\color[HTML]{1E1E1E} 0.50±0.05} & {\color[HTML]{1E1E1E} 0.50±0.05} & {\color[HTML]{1E1E1E} 0.51±0.05} & {\color[HTML]{1E1E1E} 51.25±5.02} & {\color[HTML]{1E1E1E} 50.00±4.80} & {\color[HTML]{1E1E1E} 53.00±5.06} \\
10\%                                                                & {\color[HTML]{1E1E1E} 0.60±0.05} & {\color[HTML]{1E1E1E} 0.61±0.04} & {\color[HTML]{1E1E1E} 0.61±0.05} & {\color[HTML]{1E1E1E} 60.50±4.88} & {\color[HTML]{1E1E1E} 61.75±3.93} & {\color[HTML]{1E1E1E} 62.50±4.60} \\
20\%                                                                & {\color[HTML]{1E1E1E} 0.65±0.03} & {\color[HTML]{1E1E1E} 0.70±0.04} & {\color[HTML]{1E1E1E} 0.68±0.02} & {\color[HTML]{1E1E1E} 69.25±2.92} & {\color[HTML]{1E1E1E} 71.00±3.90} & {\color[HTML]{1E1E1E} 69.00±2.89} \\
40\%                                                                & {\color[HTML]{1E1E1E} 0.79±0.04} & {\color[HTML]{1E1E1E} 0.81±0.04} & {\color[HTML]{1E1E1E} 0.80±0.03} & {\color[HTML]{1E1E1E} 80.75±3.40} & {\color[HTML]{1E1E1E} 81.75±3.52} & {\color[HTML]{1E1E1E} 81.50±3.05} \\
60\%                                                                & {\color[HTML]{1E1E1E} 0.87±0.03} & {\color[HTML]{1E1E1E} 0.87±0.03} & {\color[HTML]{1E1E1E} 0.86±0.03} & {\color[HTML]{1E1E1E} 87.70±3.48} & {\color[HTML]{1E1E1E} 87.75±3.10} & {\color[HTML]{1E1E1E} 86.50±3.00} \\
80\%                                                                & {\color[HTML]{1E1E1E} 0.89±0.03} & {\color[HTML]{1E1E1E} 0.90±0.02} & {\color[HTML]{1E1E1E} 0.89±0.02} & {\color[HTML]{1E1E1E} 89.25±3.02} & {\color[HTML]{1E1E1E} 90.75±2.17} & {\color[HTML]{1E1E1E} 89.75±0.02} \\
100\%                                                               & {\color[HTML]{1E1E1E} 0.90±0.02} & {\color[HTML]{1E1E1E} 0.91±0.02} & {\color[HTML]{1E1E1E} 0.90±0.02} & {\color[HTML]{1E1E1E} 90.55±2.05} & {\color[HTML]{1E1E1E} 91.85±1.77} & {\color[HTML]{1E1E1E} 91.00±2.32} \\ \hline
\end{tabular}%
}
\vspace{-3mm}
\end{table}
\begin{table}[t]
\caption{Performance measure of proposed method MPCS (ResNet-50 encoder) on Breast Cancer Cell Dataset}
\resizebox{\columnwidth}{!}{%
\label{tab:bisque_results_sota_compare}
\begin{tabular}{c|cccccc}
\hline
                         & \multicolumn{3}{c}{Fine-tuned}                                                                                                                 & \multicolumn{3}{c}{Linear-evaluation}                                                                                              \\ \cline{2-7} 
\multirow{-2}{*}{Method} & accuracy                                   & precision                                 & \multicolumn{1}{c|}{recall}                           & accuracy                                   & precision                                 & recall                                    \\ \hline
ST~\cite{rezaeilouyeh2016microscopic}                       & {\color[HTML]{1E1E1E} 86.00±3.00}          & {\color[HTML]{1E1E1E} -}                  & \multicolumn{1}{c|}{\textbf{1.0}}                     & {\color[HTML]{1E1E1E} -}                   & {\color[HTML]{1E1E1E} -}                  & -                                         \\
MATN~\cite{konstantinov2022multi}                     & {\color[HTML]{1E1E1E} 91.70}               & {\color[HTML]{1E1E1E} -}                  & \multicolumn{1}{c|}{-}                                & {\color[HTML]{1E1E1E} -}                   & {\color[HTML]{1E1E1E} -}                  & -                                         \\
ATN~\cite{ilse2018attention}                      & {\color[HTML]{1E1E1E} 75.50±1.60}          & {\color[HTML]{1E1E1E} 0.73±0.01}          & \multicolumn{1}{c|}{{\color[HTML]{1E1E1E} 0.73±0.04}} & -                                          & -                                         & -                                         \\ \hline
MPCS-FP                  & {\color[HTML]{1E1E1E} 98.14±2.05}          & {\color[HTML]{1E1E1E} 0.99±0.01}          & \multicolumn{1}{c|}{{\color[HTML]{1E1E1E} 0.98±0.01}} & {\color[HTML]{1E1E1E} 96.29±1.90}          & {\color[HTML]{1E1E1E} 0.97±0.01}          & {\color[HTML]{1E1E1E} 0.96±0.01}          \\
MPCS-OP                  & {\color[HTML]{1E1E1E} \textbf{98.18±1.80}} & {\color[HTML]{1E1E1E} \textbf{0.99±0.01}} & \multicolumn{1}{c|}{{\color[HTML]{1E1E1E} 0.98±0.01}} & {\color[HTML]{1E1E1E} \textbf{96.36±1.88}} & {\color[HTML]{1E1E1E} \textbf{0.97±0.01}} & {\color[HTML]{1E1E1E} \textbf{0.96±0.01}} \\
MPCS-RP                  & {\color[HTML]{1E1E1E} 98.10±2.00}          & {\color[HTML]{1E1E1E} 0.985±0.01}         & \multicolumn{1}{c|}{{\color[HTML]{1E1E1E} 0.98±0.01}} & {\color[HTML]{1E1E1E} 96.22±2.02}          & {\color[HTML]{1E1E1E} 0.965±0.01}         & {\color[HTML]{1E1E1E} 0.96±0.01}          \\ \hline
\end{tabular}%
}
\end{table}


The quantitative results and qualitative analysis from extensive experiments on three datasets validate the efficiency of learned representations for the proposed self-supervised pre-training method MPCS. Preliminary investigation on data prior also supports the hypothesis of reducing human prior in self-supervised representation learning.

\begin{figure}[!ht]
    \centering
    \includegraphics[width =0.9\linewidth]{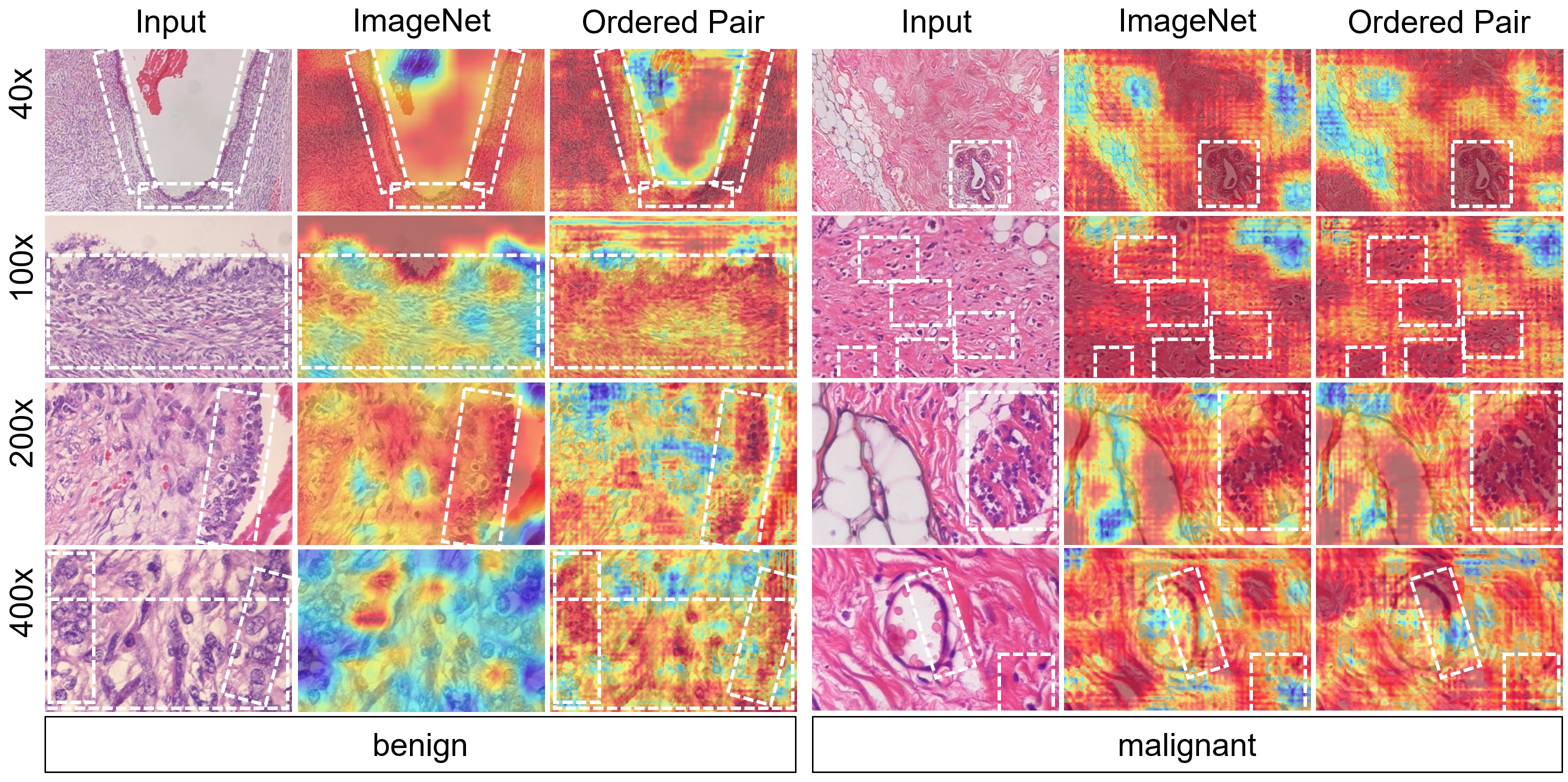}
    \caption{Comparing CAM from ImageNet and MPCS-OP on BreakHis dataset. Only OP variant CAM is shown for space constraint. Activation - red (darker) color}
    \label{fig:cam_breakhis}
    \vspace{-6mm}
\end{figure}

\begin{figure}[!ht]
    \centering
    \includegraphics[width =0.9\linewidth]{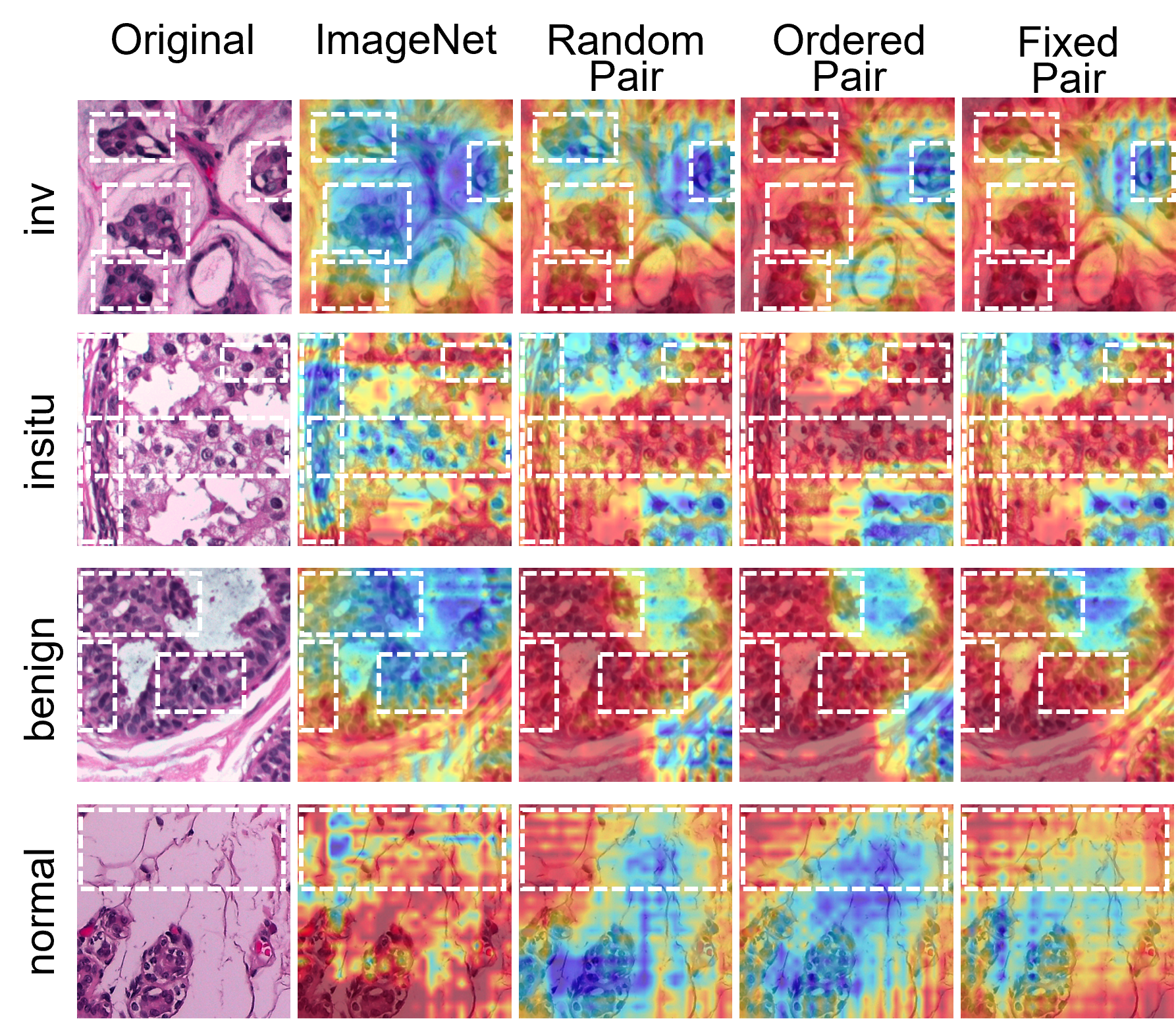}
    \caption{CAM from ImageNet and MPCS-RP, OP, FP on BACH dataset (MPCS specifically learns to prevent activation on cells present on normal class images).}
    \label{fig:cam_bach}
    \vspace{-6mm}
\end{figure}

\begin{figure}[!ht]
    \centering
    \includegraphics[width =\linewidth]{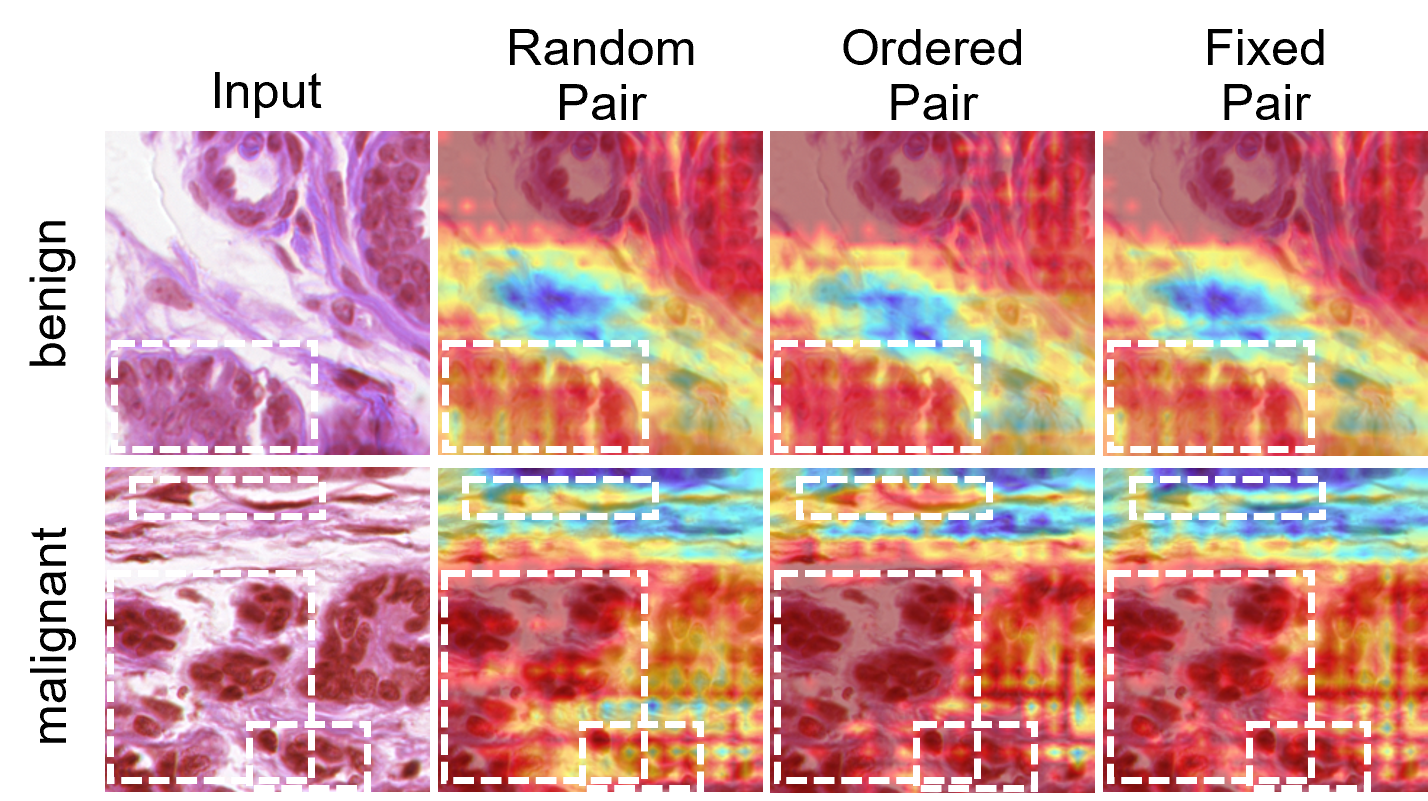}
    \caption{Comparing CAM among MPCS-RP, OP, and FP on Breast Cancer Cell dataset. Activation - red (darker) color}
    \label{fig:cam_bisque}
    \vspace{-6mm}
\end{figure}

Results on BreakHis dataset for the experiments Exp-1 to Exp-4 in Table~\ref{tab:experiments_details}  in limited labels setting (20\% labels only) are described in Table~\ref{ft_20}. It shows that all the variants (ordered pair, random pair, and fixed pair) of MPCS pre-trained models obtain significant \textit{(across magnification, p \textless 0.05)} improvement (1.55 $\sim$ 2.52)\% over the ImageNet transfer learning model for all four magnifications and results are competitive with other state-of-the-art methods that have been trained in the fully supervised setting using all labels. Further, Table~\ref{ft_80_sota} compares the performance of experiments Exp-5 to Exp-8 in Table~\ref{tab:experiments_details} in the fully supervised setting (using all labels). The MPCS pre-trained models outperform several state-of-the-art methods with (3.5 $\sim$ 8.0)\% higher accuracy in malignancy classification. Both encoder architectures, Efficient-net b2 and ResNet-50 perform consistently. The t-SNE visualization of the pre-trained ResNet-50 encoder in Figure~\ref{fig:tsne} and class activation maps (CAM)~\cite{selvaraju2017grad} in Figure~\ref{fig:cam_breakhis} show robust representation learning in the pre-tuning and fine-tuning phases, respectively. Additionally, cross-magnification evaluation is also performed as suggested in previous work~\cite{gupta2017breast} and \cite{sikaroudi2021magnification}, results support performance generalization across magnifications, and results can be found in the supplementary material.
MPCS pre-trained ResNet-50 models on the BreakHis dataset are further evaluated on two additional datasets. Results on the BACH dataset for the experiments Exp-9 to Exp-12 in Table~\ref{tab:exp_bach} are described in Table~\ref{tab:bach_results_sota_compare} and ~\ref{tab:bach_results_labelwise}. Results in Table~\ref{tab:bach_results_sota_compare} show the performance improvement on four-class classification on image-wise and patch-wise accuracy while comparing with other state-of-the-arts for the ResNet-50 encoder and other architectures. Specifically, compared with other recent work based on contrastive learning-based self-supervised method DPCL~\cite{ciga2022self}, the proposed method MPCS consistently outperforms over a varying range of label usage as shown in Figure~\ref{fig:ssl_dpcl} and obtains an improvement of 4.85\% when using 100\% labels. To compare with ImageNet pre-trained ResNet-50 encoder in identical settings, the method suggested in~\cite{vesal2018classification} is re-implemented and results are reported in Table~\ref{tab:bach_results_sota_compare} for comparison and qualitative analysis through CAM is also shown for samples from the BACH dataset in Figure~\ref{fig:cam_bach}.
Results on the third Breast Cell Cancer dataset for experiments Exp-13 to Exp-15 from Table~\ref{tab:exp_bisque} are compared with other methods in Table~\ref{tab:bisque_results_sota_compare}, shows significant improvement over other methods and obtains 98.18\% in fine-tuning and 96.36\% in linear evaluation. CAM in Figure~\ref{fig:cam_bisque} also supports the method qualitatively. Reproducible source code for all the results reported is added in supplementary content and shall be made available on GitHub. 

\subsection{Self-supervised method MPCS demonstrates label efficiency}

\begin{figure}[t]
    \centering
    \includegraphics[width =0.6\linewidth]{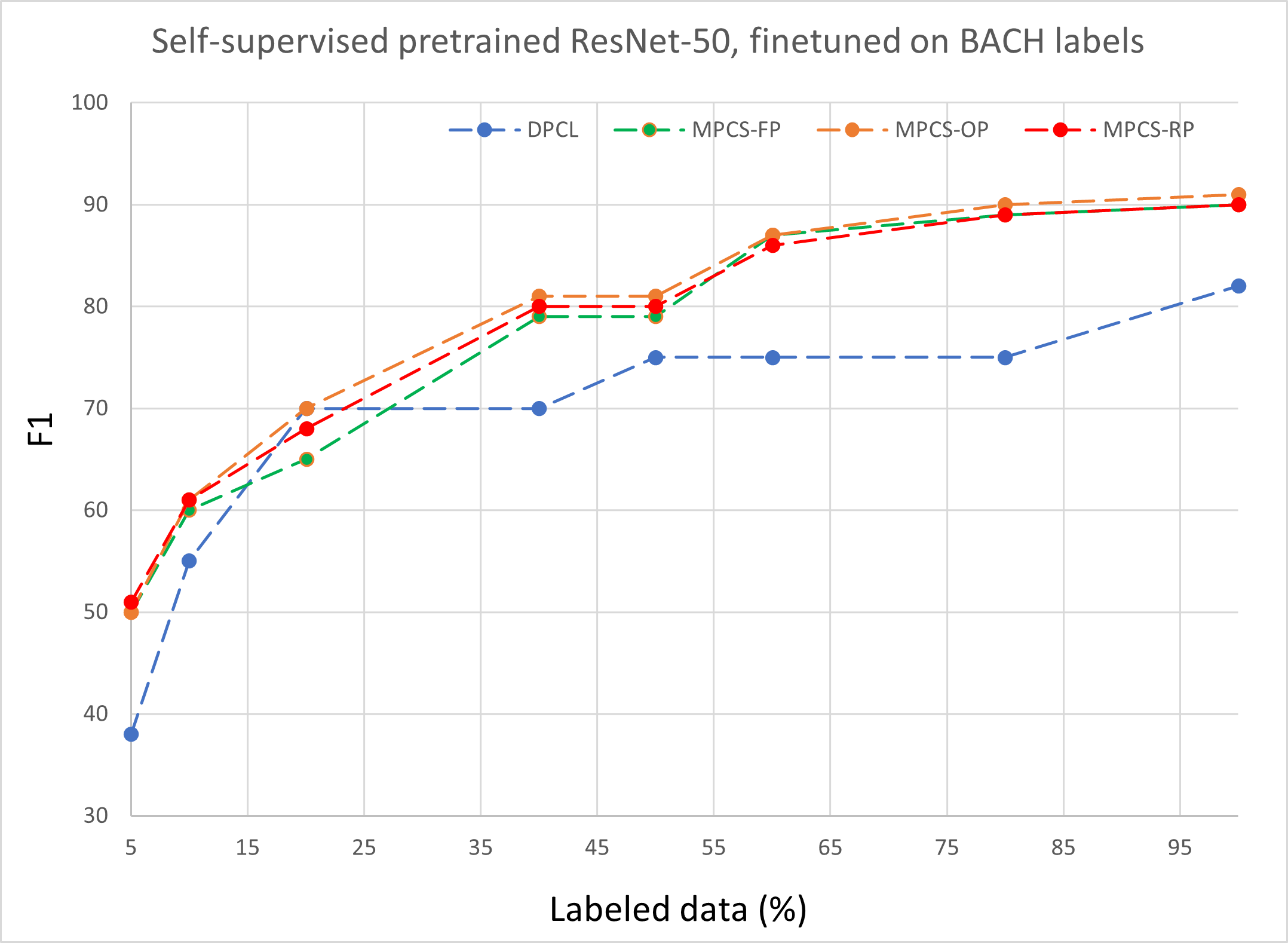}
    \caption{Comparison with DPCL~\cite{ciga2022self} for label efficiency on classification task on BACH dataset}
    \label{fig:ssl_dpcl}
    \vspace{-7mm}
\end{figure}
All three variants of the proposed method MPCS demonstrate label efficiency on downstream tasks. Results in Table~\ref{ft_20} On BreakHis dataset, MPCS fine-tuned models obtain (significant improvement p \textless 0.01) proportionally bigger margins of improvement of (2.52±0.02)\% over ImageNet pre-trained model while only 20 \% labels are used for all magnification scales. These results match the performance of other state-of-the-art methods mentioned in Table~\ref{ft_80_sota} which have been trained on 100\% labels. Following the trend, MPCS pre-trained models consistently outperform on the BACH dataset compared with the recent contrastive learning-based method DPCL~\cite{ciga2022self} in a complete range of labels from 5\% to 100\%, shown in Figure~\ref{fig:ssl_dpcl}. 

\subsection{Data prior enables self-supervision on small-scale datasets}
The proposed MPCS method enables self-supervised representation learning on the small-scale dataset by extensively using a data prior (supervision signal from data) e.g. magnification factors (40X, 100X, 200X, and 400x). It decreases the dependence on human-curated priors e.g. augmentation method choices during self-supervised pre-training.  
\subsection{MPCS learns robust self-supervised representations}
\begin{figure}[!ht]
    \centering
    \includegraphics[width =0.8\linewidth]{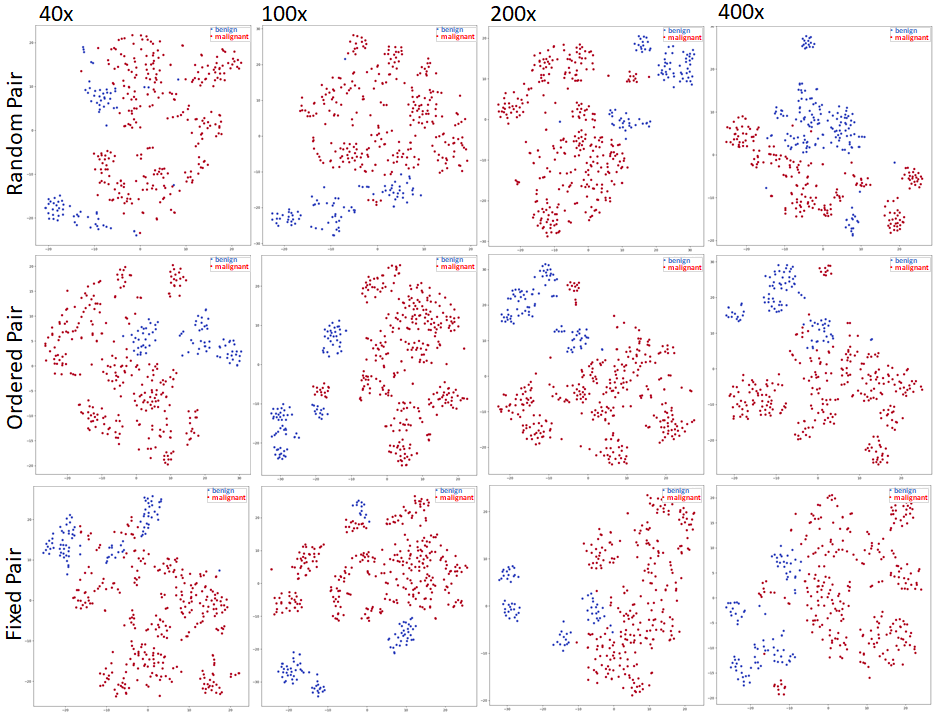}
    \caption{tsne-visualization of the features from MPCS model (on BreakHis) without fine-tuning. Blue(benign); red(malignant).  }
    \label{fig:tsne}
    \vspace{-4mm}
\end{figure}
Explicitly focusing on the robustness of learned representations, Figure~\ref{fig:tsne} strongly supports the fact that MPCS pre-trained models can capture and learn discriminative features across the classes during the self-supervised pre-training phase itself without knowing about actual human-provided labels. It is worth mentioning that data points of different classes are easily separable by either linear or non-linear boundaries for all four magnifications for all variants of the MPCS method. The CAM of fine-tuned models depicted in Figures~\ref{fig:cam_breakhis} and ~\ref{fig:cam_bach} also indicate that MPCS pre-trained models activate regions of interest (dark red colors show strong activation) more efficiently than the ImageNet pre-trained model for BreakHis and BACH datasets.  
\subsection{Preliminary support for the hypothesis  about reducing human inducted priors}
The MPCS-Ordered Pair inducts weaker human-prior in pair sampling. Thus, the MPCS method obtains one DoF for randomly choosing the first input view. In comparison, the MPCS-Fixed Pair, which inducts stronger human prior by choosing both views, 200x and 400x by human-prior, gives zero DoF to the MPCS method. The MPCS-Random Pair, in which the MPCS method obtains the highest degree of freedom since the human-prior is absent.
\begin{figure}[!ht]
    \centering
    \includegraphics[width =0.6\linewidth]{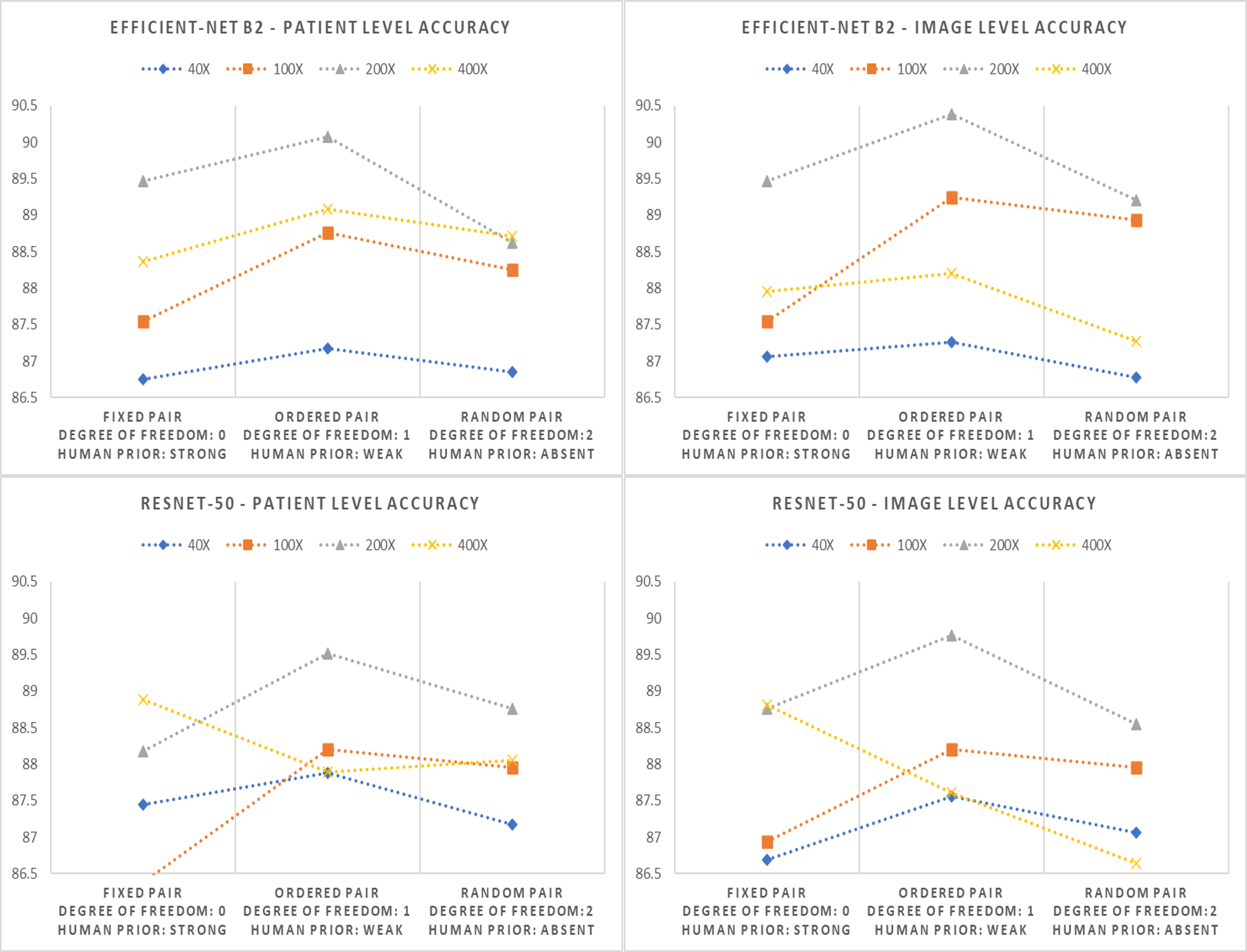}
    \caption{Comparison for human priors - Indicates weaker human-prior(moderate DoF for method) outperforms stronger and no human-prior during limited (20\% labels) data\\}
    \label{fig:degree_of_freedom_ILA_PLA}
    \vspace{-8mm}
\end{figure}
Figure~\ref{fig:degree_of_freedom_ILA_PLA} explains the trends when fewer labels are used, that both encoders that weaker human-prior based pair sampling tends to outperform over extreme cases of stronger human prior or the absence of it. However, it requires detailed explorations of different datasets and tasks. Similar patterns were also observed on the BACH dataset that the ordered pair outperforms over other two variants.

\section{Conclusions}
The novel MPCS method enables self-supervised pre-training for comparatively small-scale breast cancer microscopic image dataset BreakHis~\cite{spanhol2016dataset} for efficient representation learning by exploiting supervision signals from data.
Results on three public datasets have confirmed the excellence of the MPCS method. MPCS learns better representations by exploiting magnification priors (OP is best among all) and distinguishing among different cells in downstream tasks. It could be better to investigate on a continuous scale of magnification scales considered in this paper. In the future, we intend to investigate magnification prior to other self-supervised learning approaches focusing on redundancy reduction and Siamese networks.

{
\bibliographystyle{ieee}
\bibliography{references}
}
\pagebreak
\section{Supplementary Material}
\subsection{Self-supervised methods learn magnification invariant representations}
The MPCS methods not only outperform for magnification-specific tasks stated in Tables~\ref{ft_20} and~\ref{ft_80_sota} but representations learned through the proposed method also demonstrate a consistent edge in classification performance in cross-magnification evaluation over the ImageNet model. These experiments were conducted on the Efficient-net b2 encoder only to prevent additional computation usage. However, the ResNet-50 encoder can be benchmarked, if needed.
\begin{table}[h]
\caption{Type 1 cross magnification performance  comparison of proposed methods (Leave one magnification out on which model trained). The values represent mean performance of remaining magnifications (e.g. train on 40x and evaluated (mean) on 100x, 200x, and 400x.)}
\label{tab:results_ft80_cross_magnificaiton}
\resizebox{\columnwidth}{!}{%

\begin{tabular}{c|cccc}
\hline
                         & \multicolumn{1}{c|}{\begin{tabular}[c]{@{}c@{}}Trained \\ on 40X\end{tabular}} & \multicolumn{1}{c|}{\begin{tabular}[c]{@{}c@{}}Trained \\ on 100X\end{tabular}} & \multicolumn{1}{c|}{\begin{tabular}[c]{@{}c@{}}Trained \\ on 200X\end{tabular}} & \begin{tabular}[c]{@{}c@{}}Trained \\ on 400X\end{tabular}     \\ \cline{2-5} 
\multirow{-2}{*}{Method} & \multicolumn{4}{c}{Mean Cross-Magnification Image Level Accuracy}                                                                                                                                                                                                                                                   \\ \hline
ImageNet        & \multicolumn{1}{l|}{{\color[HTML]{444444} 79.56±11.74}}                        & \multicolumn{1}{l|}{{\color[HTML]{444444} 79.56±11.74}}                         & \multicolumn{1}{l|}{{\color[HTML]{444444} 82.97±6.77}}                          & \multicolumn{1}{l}{{\color[HTML]{444444} 84.16±4.98}}          \\
MPCS-Ordered Pair             & \multicolumn{1}{l|}{{\color[HTML]{444444} \textbf{80.99±8.91}}}                & \multicolumn{1}{l|}{{\color[HTML]{444444} \textbf{80.99±8.91}}}                 & \multicolumn{1}{l|}{{\color[HTML]{444444} 84.40±3.81}}                          & \multicolumn{1}{l}{{\color[HTML]{444444} 84.20±5.58}}          \\
MPCS-Random Pair              & \multicolumn{1}{l|}{{\color[HTML]{444444} 80.13±9.60}}                         & \multicolumn{1}{l|}{{\color[HTML]{444444} 80.13±9.60}}                          & \multicolumn{1}{l|}{{\color[HTML]{444444} \textbf{84.84±5.30}}}                 & \multicolumn{1}{l}{{\color[HTML]{444444} \textbf{84.83±5.30}}} \\ \hline
\multicolumn{1}{l|}{}    & \multicolumn{4}{c}{Mean Cross-Magnification Patient Level Accuracy}                                                                                                                                                                                                                                                 \\ \hline
ImageNet        & \multicolumn{1}{c|}{{\color[HTML]{444444} 81.01±9.59}}                         & \multicolumn{1}{c|}{{\color[HTML]{444444} 84.38±5.78}}                          & \multicolumn{1}{c|}{{\color[HTML]{444444} 81.45±6.89}}                          & {\color[HTML]{444444} 83.31±7.31}                              \\
MPCS-Ordered Pair             & \multicolumn{1}{c|}{{\color[HTML]{444444} \textbf{81.49±6.96}}}                & \multicolumn{1}{c|}{{\color[HTML]{444444} 84.02±5.78}}                          & \multicolumn{1}{c|}{{\color[HTML]{444444} 82.19±3.83}}                          & {\color[HTML]{444444} 83.10±7.44}                              \\
MPCS-Random Pair              & \multicolumn{1}{c|}{{\color[HTML]{444444} 81.20±7.09}}                         & \multicolumn{1}{c|}{{\color[HTML]{444444} \textbf{85.05±7.15}}}                 & \multicolumn{1}{c|}{{\color[HTML]{444444} \textbf{82.97±5.84}}}                 & {\color[HTML]{444444} \textbf{83.76±6.63}}                     \\ \hline
\end{tabular}%
}
\label{x_mag_type1}
\end{table} 
\begin{table}[h]
\caption{Type 2 cross magnification performance  comparison of proposed methods (select one magnification in on which model was not trained). The values represent mean performance of a  magnification whereas trained on other magnifications (e.g. evaluated on 40x and trained on 100x, 200x, and 400x.}
\label{tab:results_ft80_cross_magnificaiton_type2}
\resizebox{\columnwidth}{!}{%
\begin{tabular}{c|llll}
\hline
                         & \multicolumn{1}{c|}{\begin{tabular}[c]{@{}c@{}}Evaluated \\ on 40X\end{tabular}} & \multicolumn{1}{c|}{\begin{tabular}[c]{@{}c@{}}Evaluated \\ on 100X\end{tabular}} & \multicolumn{1}{c|}{\begin{tabular}[c]{@{}c@{}}Evaluated \\ on 200X\end{tabular}} & \multicolumn{1}{c}{\begin{tabular}[c]{@{}c@{}}Evaluated \\ on 400X\end{tabular}} \\ \cline{2-5} 
\multirow{-2}{*}{Method} & \multicolumn{4}{c}{Mean Cross-Magnification Image Level Accuracy}                                                                                                                                                                                                                                                                           \\ \hline
ImageNet        & \multicolumn{1}{l|}{{\color[HTML]{444444} 84.33 ±   4.32}}                       & \multicolumn{1}{l|}{{\color[HTML]{444444} 83.65 ±   6.07}}                        & \multicolumn{1}{l|}{{\color[HTML]{444444} 84.31 ±   8.93}}                        & {\color[HTML]{444444} 78.86 ±   10.62}                                           \\
MPCS-Ordered Pair             & \multicolumn{1}{l|}{{\color[HTML]{444444} 85.35 ±   4.29}}                       & \multicolumn{1}{l|}{{\color[HTML]{444444} 84.56 ±   6.06}}                        & \multicolumn{1}{l|}{{\color[HTML]{444444} \textbf{85.28 ±   8.20}}}               & {\color[HTML]{444444} 78.98 ±   7.82}                                            \\
MPCS-Random Pair              & \multicolumn{1}{l|}{{\color[HTML]{444444} \textbf{86.55 ±   5.10}}}              & \multicolumn{1}{l|}{{\color[HTML]{444444} \textbf{84.82 ±   5.54}}}               & \multicolumn{1}{l|}{{\color[HTML]{444444} 84.99 ±   7.79}}                        & {\color[HTML]{444444} \textbf{79.17 ±   9.60}}                                   \\ \hline
\multicolumn{1}{l|}{}    & \multicolumn{4}{c}{Mean Cross-Magnification Patient Level Accuracy}                                                                                                                                                                                                                                                                         \\ \hline
ImageNet        & \multicolumn{1}{l|}{{\color[HTML]{444444} 81.71 ±   6.25}}                       & \multicolumn{1}{l|}{{\color[HTML]{444444} 83.76 ±   6.82}}                        & \multicolumn{1}{l|}{{\color[HTML]{444444} 84.35 ±   8.40}}                        & {\color[HTML]{444444} \textbf{80.33 ±   8.10}}                                   \\
MPCS-Ordered Pair             & \multicolumn{1}{l|}{{\color[HTML]{444444} 81.97 ±   6.25}}                       & \multicolumn{1}{l|}{{\color[HTML]{444444} 84.54 ±   6.52}}                        & \multicolumn{1}{l|}{{\color[HTML]{444444} 84.63 ±   7.80}}                        & {\color[HTML]{444444} 79.51 ±   5.20}                                            \\
MPCS-Random Pair              & \multicolumn{1}{l|}{{\color[HTML]{444444} \textbf{83.24 ±   6.89}}}              & \multicolumn{1}{l|}{{\color[HTML]{444444} \textbf{84.98 ±   6.12}}}               & \multicolumn{1}{l|}{{\color[HTML]{444444} \textbf{84.84 ±   7.04}}}               & {\color[HTML]{444444} 79.92 ±   6.53}                                            \\ \hline
\end{tabular} %
}
\label{x_mag_type2}
\end{table}
Table~\ref{x_mag_type1} shows (type-1 mean cross magnification evaluation) mean cross-magnification accuracy where the model is evaluated on other magnifications except the magnification on which the model was trained.
The MPCS-Order Pair methods outperform the ImageNet model and other methods with a mean cross-magnification ILA of 80.99\% and PLA of 81.49\% when the model was trained on 40x magnification and evaluated on 100x, 200x, and 400x.
Whereas MPCS-Random Pair outperforms with mean cross-magnification ILA 84.84\% and PLA 82.97\% when trained on 200x and ILA 84.83\% and PLA 83.76\% when trained on 400x.
For 100x, MPCS-Ordered Pair obtains ILA 80.99\%, and MPCS-Random Pair obtains PLA 85.05\%.
Further, Table~\ref{x_mag_type2} evaluates the mean performance of models trained on other magnifications except on which evaluation is performed (type-2 mean cross magnification evaluation).
Interestingly, type-2 cross magnification evaluation also shows similar trends except in 400x, in which the ImageNet model obtained high PLA performance.
Empirical analysis on type-1 and type-2 cross magnification suggests that MPCS self-supervised pre-trained models perform better than the ImageNet model by learning magnification invariant representations.
\subsection{Additional ablation results on using 80\%, 60\%, and 40\% of labels of the training set on BreakHis dataset}
\begin{figure}[]
    \centering
    \includegraphics[width =0.9\linewidth]{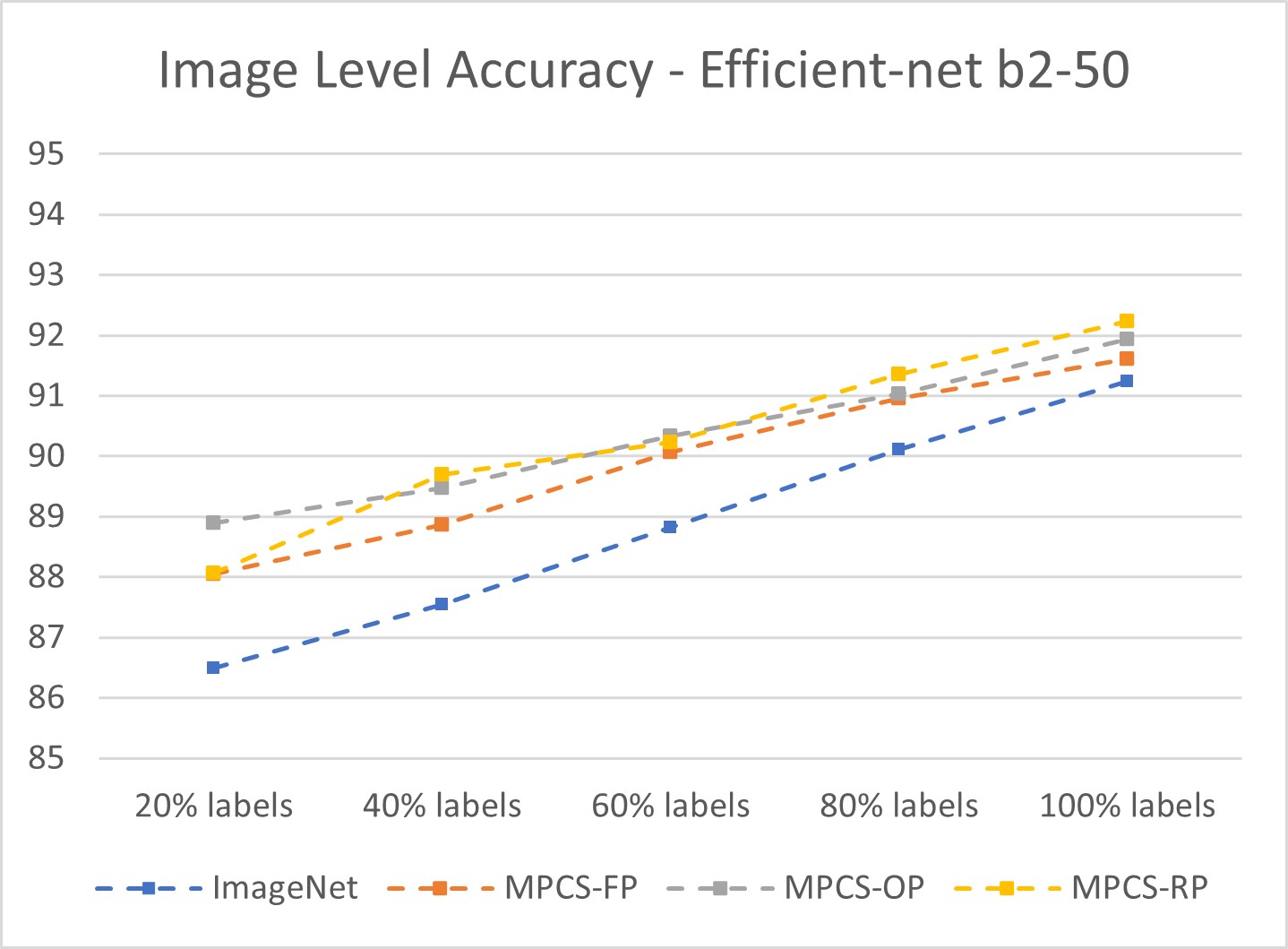}
    \caption{Performance comparison (ILA accuracy, Efficient-net b2 model) for MPCS pre-trained models with ImageNet pre-trained model over range of labels used.}
    \label{fig:effnet_ila_compare}
\end{figure}

\begin{figure}[]
    \centering
    \includegraphics[width =0.9\linewidth]{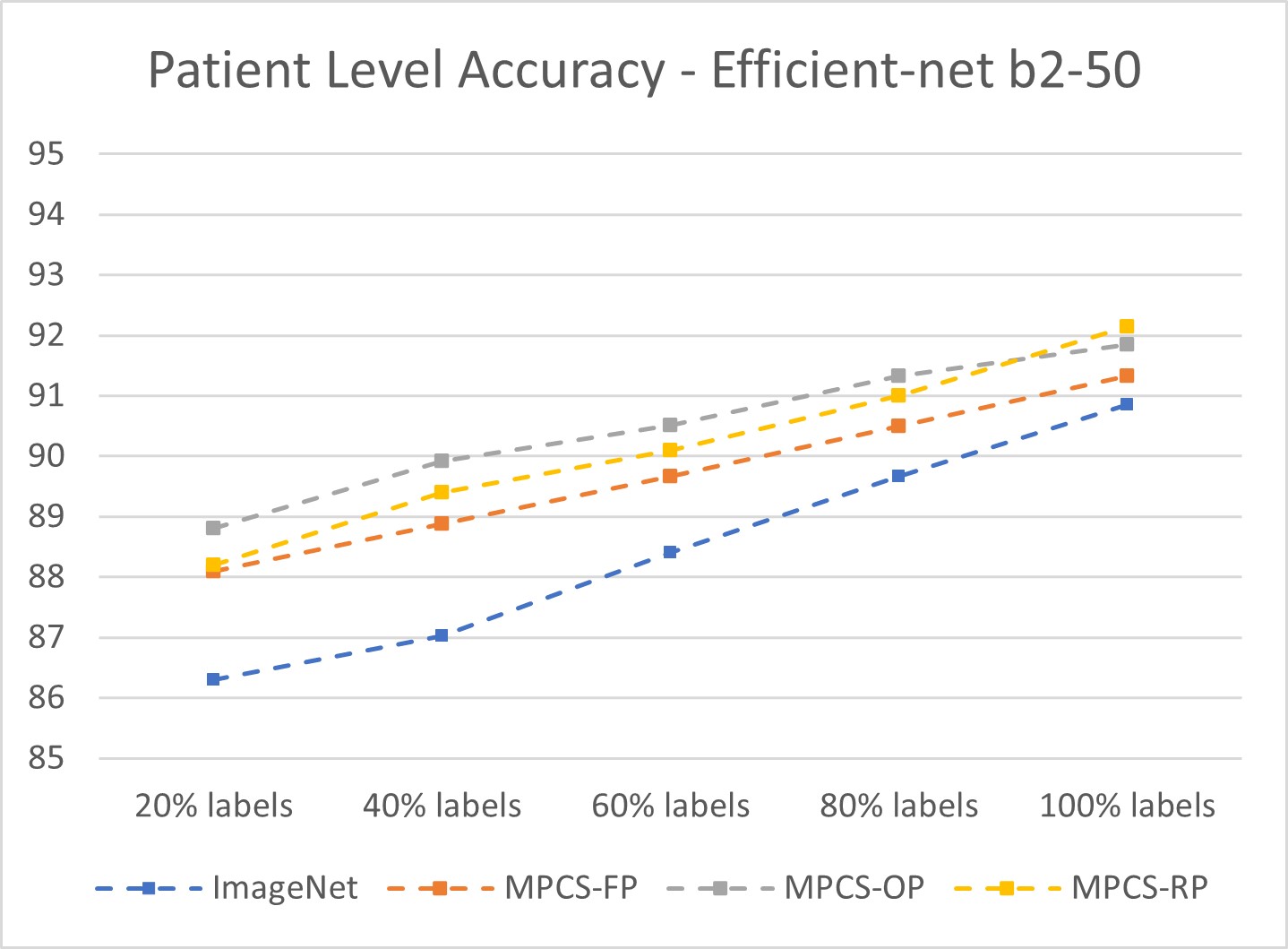}
    \caption{Performance comparison (PLA accuracy, Efficient-net b2 model) for MPCS pre-trained models with ImageNet pre-trained model over range of labels used.}
    \label{fig:effnet_pla_compare}
\end{figure}

\begin{figure}[t]
    \centering
    \includegraphics[width =0.9\linewidth]{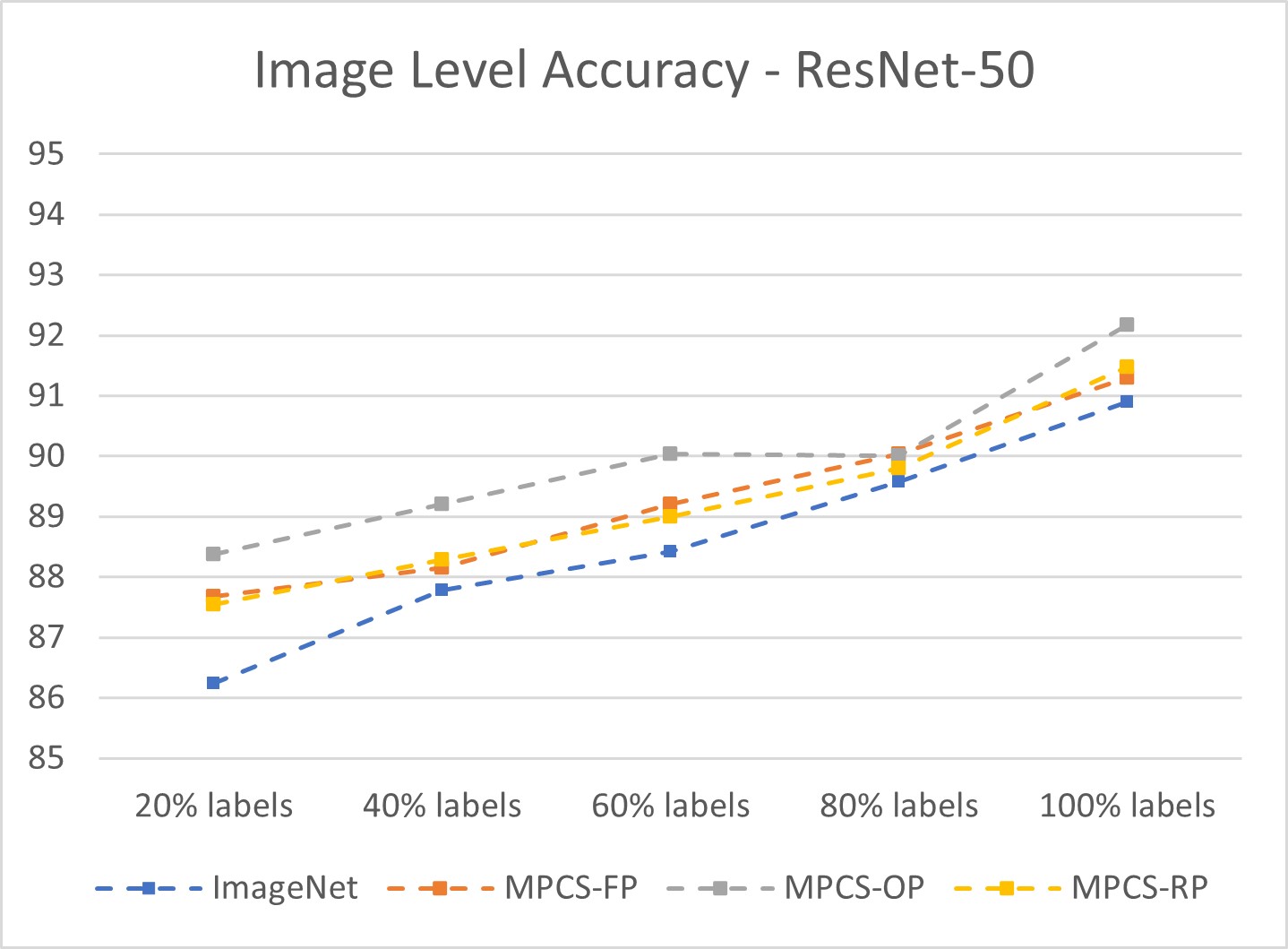}
    \caption{Performance comparison (ILA accuracy, ResNet-50 model) for MPCS pre-trained models with ImageNet pre-trained model over range of labels used.}
    \label{fig:resnet_ila_compare}
\end{figure}

\begin{figure}[t]
    \centering
    \includegraphics[width =0.9\linewidth]{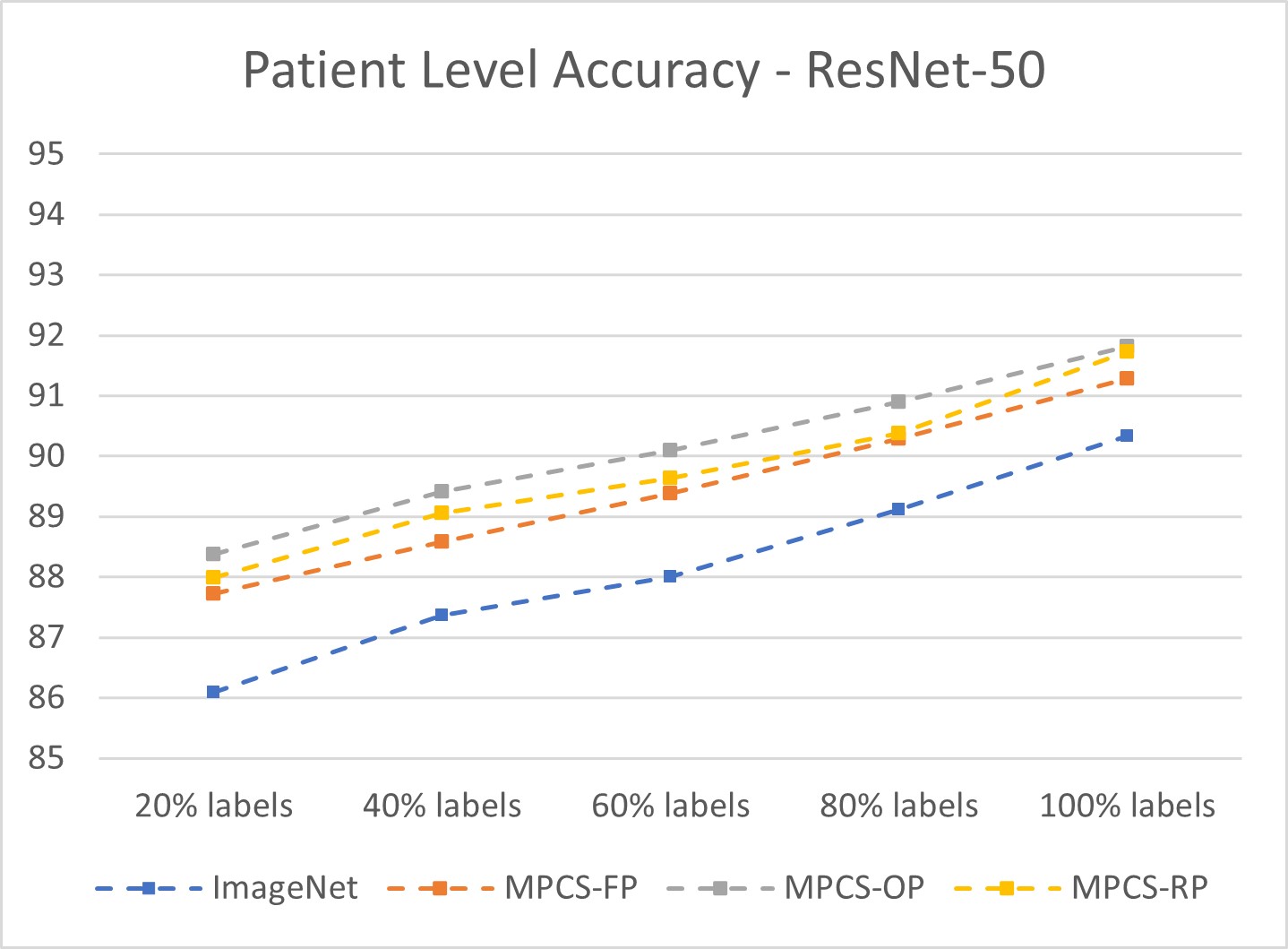}
    \caption{Performance comparison (PLA accuracy, ResNet-50 model) for MPCS pre-trained models with ImageNet pre-trained model over range of labels used.}
    \label{fig:resnet_pla_compare}
\end{figure}

This section describes the extended ablation study on using labels in an incremental manner. The main section of the results describes and compares all three variants of the MPCS method with ImageNet pre-trained models when fine-tuned on 20\% and 100\% labels of the training set are used.
\begin{table*}[h]
\centering
\caption{Performance evaluation of the proposed methods in limited labelled data setting when fine-tuning on 80\% labels of train set.}
\label{tab:breakhis_80_label_mpcs}
\begin{adjustbox}{max width=\textwidth}
\begin{tabular}{c|cccc|c|cccc|c}
\hline
                         & \multicolumn{4}{c|}{Patient Level Accuracy (RR)}                                                                                                                                       &                        & \multicolumn{4}{c|}{Image Level Accuracy}                                                                        &                        \\ \cline{2-5} \cline{7-10}
\multirow{-2}{*}{Method} & \multicolumn{1}{c|}{40X}        & \multicolumn{1}{c|}{100X}                               & \multicolumn{1}{c|}{200X}                              & 400X                              & \multirow{-2}{*}{Mean} & \multicolumn{1}{c|}{40X}        & \multicolumn{1}{c|}{100X}       & \multicolumn{1}{c|}{200X}       & 400X       & \multirow{-2}{*}{Mean} \\ \hline
ImageNet (Eff-net b2)    & \multicolumn{1}{c|}{90.48±3.45} & \multicolumn{1}{c|}{90.53±3.60}                         & \multicolumn{1}{c|}{90.36±4.67}                        & 87.30±3.88                        & 89.66±3.90             & \multicolumn{1}{c|}{91.32±3.48} & \multicolumn{1}{c|}{91.46±3.40} & \multicolumn{1}{c|}{90.34±3.55} & 87.32±4.76 & 90.11±3.79             \\
MPCS-FP (Eff-net b2)     & \multicolumn{1}{c|}{91.26±3.64} & \multicolumn{1}{c|}{91.54±3.68}                         & \multicolumn{1}{c|}{90.84±2.66}                        & 88.32±3.23                        & 90.49±3.30             & \multicolumn{1}{c|}{91.83±3.48} & \multicolumn{1}{c|}{92.34±3.20} & \multicolumn{1}{c|}{91.33±2.45} & 88.30±3.50 & 90.95±3.15             \\
MPCS-OP (Eff-net b2)     & \multicolumn{1}{c|}{92.05±3.17} & \multicolumn{1}{c|}{92.73±2.45}                         & \multicolumn{1}{c|}{{\color[HTML]{1E1E1E} 91.54±2.80}} & {\color[HTML]{1E1E1E} 88.98±3.42} & 91.33±2.96             & \multicolumn{1}{c|}{91.67±3.53} & \multicolumn{1}{c|}{92.45±2.80} & \multicolumn{1}{c|}{91.45±3.60} & 88.55±3.40 & 91.03±3.33             \\
MPCS-RP (Eff-net b2)     & \multicolumn{1}{c|}{92.16±4.18} & \multicolumn{1}{c|}{{\color[HTML]{1E1E1E} 91.51± 3.18}} & \multicolumn{1}{c|}{{\color[HTML]{1E1E1E} 91.13±2.81}} & {\color[HTML]{1E1E1E} 89.16±2.43} & 91.00±3.15             & \multicolumn{1}{c|}{92.25±3.61} & \multicolumn{1}{c|}{92.48±3.20} & \multicolumn{1}{c|}{92.01±3.49} & 88.71±3.15 & 91.36±3.79             \\
ImageNet (RN-50)         & \multicolumn{1}{c|}{90.16±3.40} & \multicolumn{1}{c|}{90.01±4.1}                          & \multicolumn{1}{c|}{89.32±3.54}                        & 87.00±4.45                        & 89.12±3.87             & \multicolumn{1}{c|}{90.53±5.12} & \multicolumn{1}{c|}{91.03±3.23} & \multicolumn{1}{c|}{90.31±3.59} & 86.48±4.92 & 89.58±4.21             \\
MPCS-FP (RN-50)          & \multicolumn{1}{c|}{90.44±3.43} & \multicolumn{1}{c|}{91.32±2.54}                         & \multicolumn{1}{c|}{90.43±3.32}                        & 88.95±2.10                        & 90.29±2.85             & \multicolumn{1}{c|}{90.14±3.48} & \multicolumn{1}{c|}{91.00±3.88} & \multicolumn{1}{c|}{90.71±3.24} & 88.30±2.46 & 90.04±3.27             \\
MPCS-OP (RN-50)          & \multicolumn{1}{c|}{92.00±2.42} & \multicolumn{1}{c|}{92.16±3.08}                         & \multicolumn{1}{c|}{{\color[HTML]{1E1E1E} 91.15±2.88}} & {\color[HTML]{1E1E1E} 88.30±3.30} & 90.90±2.92             & \multicolumn{1}{c|}{92.16±2.80} & \multicolumn{1}{c|}{92.15±2.89} & \multicolumn{1}{c|}{91.05±2.43} & 88.50±2.45 & 90.00±2.64             \\
MPCS-RP (RN-50)          & \multicolumn{1}{c|}{91.02±3.56} & \multicolumn{1}{c|}{{\color[HTML]{1E1E1E} 91.20± 3.28}} & \multicolumn{1}{c|}{{\color[HTML]{1E1E1E} 91.10±3.88}} & {\color[HTML]{1E1E1E} 88.20±3.17} & 90.38±2.92             & \multicolumn{1}{c|}{91.02±2.44} & \multicolumn{1}{c|}{91.22±3.13} & \multicolumn{1}{c|}{90.02±3.65} & 87.06±3.84 & 89.80±3.27             \\ \hline
\end{tabular}
\end{adjustbox}
\end{table*}
\begin{table*}[h]
\centering
\caption{Performance evaluation of the proposed methods in limited labelled data setting when fine-tuning on 60\% labels of train set.}
\label{tab:breakhis_60_label_mpcs}
\begin{adjustbox}{max width=\textwidth}
\begin{tabular}{c|cccc|c|cccc|c}
\hline
                         & \multicolumn{4}{c|}{Patient Level Accuracy (RR)}                                                                                                                                       &                                   & \multicolumn{4}{c|}{Image Level Accuracy}                                                                        &                        \\ \cline{2-5} \cline{7-10}
\multirow{-2}{*}{Method} & \multicolumn{1}{c|}{40X}        & \multicolumn{1}{c|}{100X}                               & \multicolumn{1}{c|}{200X}                              & 400X                              & \multirow{-2}{*}{Mean}            & \multicolumn{1}{c|}{40X}        & \multicolumn{1}{c|}{100X}       & \multicolumn{1}{c|}{200X}       & 400X       & \multirow{-2}{*}{Mean} \\ \hline
ImageNet (Eff-net b2)    & \multicolumn{1}{c|}{89.28±3.56} & \multicolumn{1}{c|}{89.03±3.40}                         & \multicolumn{1}{c|}{89.22±3.12}                        & 86.10±3.32                        & 88.40±3.35                        & \multicolumn{1}{c|}{90.02±3.54} & \multicolumn{1}{c|}{90.16±3.53} & \multicolumn{1}{c|}{89.11±3.21} & 86.00±3.52 & 88.82±3.45             \\
MPCS-FP (Eff-net b2)     & \multicolumn{1}{c|}{90.22±3.87} & \multicolumn{1}{c|}{90.45±3.20}                         & \multicolumn{1}{c|}{89.72±2.70}                        & 88.32±3.21                        & 89.67±3.25                        & \multicolumn{1}{c|}{90.56±3.32} & \multicolumn{1}{c|}{91.28±3.43} & \multicolumn{1}{c|}{90.20±2.55} & 88.20±3.40 & 90.06±3.30             \\
MPCS-OP (Eff-net b2)     & \multicolumn{1}{c|}{91.10±3.21} & \multicolumn{1}{c|}{91.54±2.66}                         & \multicolumn{1}{c|}{{\color[HTML]{1E1E1E} 90.44±2.65}} & {\color[HTML]{1E1E1E} 88.96±3.55} & 90.51±3.02                        & \multicolumn{1}{c|}{90.50±3.22} & \multicolumn{1}{c|}{91.74±2.66} & \multicolumn{1}{c|}{90.63±3.41} & 88.49±3.91 & 90.34±3.30             \\
MPCS-RP (Eff-net b2)     & \multicolumn{1}{c|}{91.00±4.18} & \multicolumn{1}{c|}{{\color[HTML]{1E1E1E} 90.30± 3.22}} & \multicolumn{1}{c|}{{\color[HTML]{1E1E1E} 90.10±2.61}} & {\color[HTML]{1E1E1E} 89.00±2.55} & 90.10±3.14                        & \multicolumn{1}{c|}{91.15±3.43} & \multicolumn{1}{c|}{91.26±3.11} & \multicolumn{1}{c|}{91.01±3.26} & 87.53±3.02 & 90.23±3.21             \\
ImageNet (RN-50)         & \multicolumn{1}{c|}{88.90±3.44} & \multicolumn{1}{c|}{89.05±3.3}                          & \multicolumn{1}{c|}{88.04±3.23}                        & 86.04±3.24                        & 88.00±3.30                        & \multicolumn{1}{c|}{89.01±4.10} & \multicolumn{1}{c|}{89.65±3.55} & \multicolumn{1}{c|}{89.02±3.44} & 86.00±3.12 & 88.42±3.55             \\
MPCS-FP (RN-50)          & \multicolumn{1}{c|}{89.23±3.21} & \multicolumn{1}{c|}{90.11±2.32}                         & \multicolumn{1}{c|}{89.30±3.12}                        & 88.90±3.15                        & 89.39±2.95                        & \multicolumn{1}{c|}{89.05±3.33} & \multicolumn{1}{c|}{90.05±3.55} & \multicolumn{1}{c|}{89.56±3.65} & 88.21±3.56 & 89.21±3.52             \\
MPCS-OP (RN-50)          & \multicolumn{1}{c|}{91.02±2.54} & \multicolumn{1}{c|}{91.08±3.18}                         & \multicolumn{1}{c|}{{\color[HTML]{1E1E1E} 90.05±2.78}} & {\color[HTML]{1E1E1E} 88.21±3.22} & {\color[HTML]{1E1E1E} 90.09±3.22} & \multicolumn{1}{c|}{91.05±3.66} & \multicolumn{1}{c|}{91.01±2.70} & \multicolumn{1}{c|}{90.10±2.32} & 88.00±2.60 & 90.04±2.82             \\
MPCS-RP (RN-50)          & \multicolumn{1}{c|}{90.01±3.22} & \multicolumn{1}{c|}{{\color[HTML]{1E1E1E} 90.10± 3.28}} & \multicolumn{1}{c|}{{\color[HTML]{1E1E1E} 90.21±3.56}} & {\color[HTML]{1E1E1E} 88.22±3.20} & 89.64±3.32                        & \multicolumn{1}{c|}{90.05±2.32} & \multicolumn{1}{c|}{90.02±3.16} & \multicolumn{1}{c|}{89.01±3.55} & 86.90±3.21 & 89.00±3.06             \\ \hline
\end{tabular}
\end{adjustbox}
\end{table*}
\begin{table*}[h!t]
\centering
\caption{Performance evaluation of the proposed methods in limited labelled data setting when fine-tuning on 40\% labels of train set.}
\label{tab:breakhis_40_label_mpcs}
\begin{adjustbox}{max width=\textwidth}
\begin{tabular}{c|cccc|c|cccc|c}
\hline
                         & \multicolumn{4}{c|}{Patient Level Accuracy (RR)}                                                                                                                                       &                        & \multicolumn{4}{c|}{Image Level Accuracy}                                                                        &                        \\ \cline{2-5} \cline{7-10}
\multirow{-2}{*}{Method} & \multicolumn{1}{c|}{40X}        & \multicolumn{1}{c|}{100X}                               & \multicolumn{1}{c|}{200X}                              & 400X                              & \multirow{-2}{*}{Mean} & \multicolumn{1}{c|}{40X}        & \multicolumn{1}{c|}{100X}       & \multicolumn{1}{c|}{200X}       & 400X       & \multirow{-2}{*}{Mean} \\ \hline
ImageNet (Eff-net b2)    & \multicolumn{1}{c|}{87.45±3.04} & \multicolumn{1}{c|}{88.00±3.22}                         & \multicolumn{1}{c|}{87.52±3.56}                        & 85.95±3.11                        & 87.03±3.23             & \multicolumn{1}{c|}{88.05±3.11} & \multicolumn{1}{c|}{89.14±3.22} & \multicolumn{1}{c|}{88.00±3.19} & 85.02±3.00 & 87.55±3.13             \\
MPCS-FP (Eff-net b2)     & \multicolumn{1}{c|}{88.66±3.62} & \multicolumn{1}{c|}{88.90±3.10}                         & \multicolumn{1}{c|}{89.66±2.03}                        & 88.30±3.12                        & 88.88±2.98             & \multicolumn{1}{c|}{88.02±3.72} & \multicolumn{1}{c|}{89.28±3.04} & \multicolumn{1}{c|}{90.00±2.16} & 88.15±3.06 & 88.86±3.00             \\
MPCS-OP (Eff-net b2)     & \multicolumn{1}{c|}{89.75±3.56} & \multicolumn{1}{c|}{90.63±2.66}                         & \multicolumn{1}{c|}{{\color[HTML]{1E1E1E} 90.32±3.11}} & {\color[HTML]{1E1E1E} 88.96±3.55} & 89.92±3.22             & \multicolumn{1}{c|}{88.82±3.22} & \multicolumn{1}{c|}{90.04±2.89} & \multicolumn{1}{c|}{90.60±3.43} & 88.40±3.89 & 89.47±3.36             \\
MPCS-RP (Eff-net b2)     & \multicolumn{1}{c|}{89.05±3.27} & \multicolumn{1}{c|}{{\color[HTML]{1E1E1E} 90.05± 3.42}} & \multicolumn{1}{c|}{{\color[HTML]{1E1E1E} 89.80±2.82}} & {\color[HTML]{1E1E1E} 88.70±2.65} & 89.40±3.04             & \multicolumn{1}{c|}{90.25±3.21} & \multicolumn{1}{c|}{91.00±3.36} & \multicolumn{1}{c|}{90.05±3.11} & 87.40±3.45 & 89.70±2.53             \\
ImageNet (RN-50)         & \multicolumn{1}{c|}{88.01±3.65} & \multicolumn{1}{c|}{88.50±3.43}                         & \multicolumn{1}{c|}{87.09±3.62}                        & 85.88±3.01                        & 87.37±3.43             & \multicolumn{1}{c|}{88.10±3.23} & \multicolumn{1}{c|}{88.60±3.53} & \multicolumn{1}{c|}{88.02±3.00} & 85.60±3.22 & 87.78±3.25             \\
MPCS-FP (RN-50)          & \multicolumn{1}{c|}{88.20±3.61} & \multicolumn{1}{c|}{88.17±2.62}                         & \multicolumn{1}{c|}{89.00±3.81}                        & 88.90±3.56                        & 88.58±3.4              & \multicolumn{1}{c|}{87.90±3.25} & \multicolumn{1}{c|}{88.10±3.21} & \multicolumn{1}{c|}{88.50±3.61} & 88.10±3.56 & 88.15±3.41             \\
MPCS-OP (RN-50)          & \multicolumn{1}{c|}{89.42±2.13} & \multicolumn{1}{c|}{90.28±3.64}                         & \multicolumn{1}{c|}{{\color[HTML]{1E1E1E} 89.92±2.18}} & {\color[HTML]{1E1E1E} 88.02±3.22} & 89.41±2.79             & \multicolumn{1}{c|}{89.10±3.05} & \multicolumn{1}{c|}{90.00±2.85} & \multicolumn{1}{c|}{89.90±2.52} & 87.80±2.65 & 89.20±2.78             \\
MPCS-RP (RN-50)          & \multicolumn{1}{c|}{89.01±3.32} & \multicolumn{1}{c|}{{\color[HTML]{1E1E1E} 89.10± 3.82}} & \multicolumn{1}{c|}{{\color[HTML]{1E1E1E} 90.00±3.46}} & {\color[HTML]{1E1E1E} 88.15±3.26} & 89.06±3.47             & \multicolumn{1}{c|}{88.10±2.35} & \multicolumn{1}{c|}{89.30±3.20} & \multicolumn{1}{c|}{88.90±3.05} & 86.85±3.22 & 88.29±2.96             \\ \hline
\end{tabular}
\end{adjustbox}
\end{table*}
To continue the trend for completeness of analysis, this section adds the results for the same setting considering 40\%, 60\%, and 80\% label utilization in fine-tuning, results described in Tables~\ref{tab:breakhis_80_label_mpcs}, ~\ref{tab:breakhis_60_label_mpcs}, and~\ref{tab:breakhis_40_label_mpcs}, respectively. The most important observation is that MPCS methods consistently outperform the ImageNet model over the range of labels provided and specifically, the ordered pair method remains best performing in largely. Figure~\ref{fig:effnet_ila_compare} and \ref{fig:effnet_pla_compare} shows the comparisons for Efficient-net b2 encoder for ILA and PLA accuracy. Similarly, Figure~\ref{fig:resnet_ila_compare} and \ref{fig:resnet_pla_compare} shows the comparisons for ResNet-50 encoder for ILA and PLA accuracy. A common trend is evident that MPCS methods based models consistently performs better than ImageNet based model for entire range of labels.

It clearly shows that self-supervised learned representations improve fine-tuning task performance overall range of available labels, similar to the trend observed on the BACH dataset. Besides being able to obtain relatively higher accuracy on limited label settings, more label additions are largely beneficial to self-supervised pre-trained models than ImageNet pre-trained models.

\subsection{Experimentation statistics}
\begin{table*}[t]
\centering
\caption{Experimentation statistics for proposed method MPCS}
\label{tab:exp_stats}
\begin{tabular}{cc|c}
\hline
\multicolumn{1}{c|}{Dataset}                    & Experiment Training Type          & No. of Experiments         \\ \hline
\multicolumn{1}{c|}{}                           & SSL pretrain - Efficient-net b2   & 15                         \\
\multicolumn{1}{c|}{}                           & SSL pretrain - ResNet-50          & 18                         \\
\multicolumn{1}{c|}{}                           & Downstream task- Efficient-net b2 & 400                        \\
\multicolumn{1}{c|}{\multirow{-4}{*}{BreakHis}} & Downstream task- ResNet-50        & {\color[HTML]{1E1E1E} 400} \\ \hline
\multicolumn{1}{c|}{BACH}                       & Downstream task- ResNet-50        & 140                        \\ \hline
\multicolumn{1}{c|}{Breast Cell Cancer Dataset} & Downstream task- ResNet-50        & 30                         \\ \hline
\multicolumn{2}{c|}{\textbf{Total}}                                                 & \textbf{1003}              \\ \hline
\end{tabular} 
\end{table*}
An extensive experimentation strategy was designed, and experiments were performed To evaluate all the variants of the proposed self-supervised pre-training method MPCS. Specifically, 15 pre-training experiments for Efficient-net b2 and 18 pre-training experiments for ResNet-50 were performed on the BreakHis dataset. Further learned representations from pre-training models are evaluated by 800 downstream task training experiments on the BreakHis dataset covering all four magnifications (40x, 100x, 200x, and 400x), 5-cross folds, and on a wide range of labels (5\% to 100\% train set labels). One hundred forty downstream task training experiments were performed on the BACH dataset using BreakHis MPCS pre-trained ResNet-50 models. Finally, 30 downstream tasks experiments were performed for the Breast Cell Cancer Dataset using ResNet-50 pre-trained models covering fine-tuning and linear evaluation. In this way, 1003 experiments are performed in the current work. Details are mentioned in Table~\ref{tab:exp_stats}.

\end{document}